\documentclass{aa}
\makeatletter
\renewcommand*\aa@pageof{, page \thepage{} of \pageref*{LastPage}}
\makeatother
\usepackage[titletoc,title]{appendix}
\usepackage{graphicx}
\usepackage{float}
\usepackage{txfonts}
\usepackage{color}
\newcommand{\LCDM}{$\Lambda$CDM}

\begin{document}

   \title{Incorporating baryon-driven contraction of dark matter halos in rotation curve fits}
\titlerunning{Baryonic contraction of DM halos}
\authorrunning{P. Li et al.}

   \author{Pengfei Li
          \inst{1}\fnmsep\thanks{Humboldt fellow}
          \and
          Stacy S. McGaugh\inst{2}
          \and
          Federico Lelli\inst{3}
          \and
          James M. Schombert\inst{4}
          \and
          Marcel S. Pawlowski\inst{1}}

   \institute{Leibniz-Institute for Astrophysics,
              An der Sternwarte 16, 14482 Potsdam, Germany\\
              \email{pli@aip.de; PengfeiLi0606@gmail.com}
              \and
              Department of Astronomy, Case Western Reserve University, 10900 Euclid Avenue, Cleveland, OH 44106, USA
              \and
              INAF – Arcetri Astrophysical Observatory, Largo Enrico Fermi 5, I-50125, Firenze, Italy
              \and
              Department of Physics, University of Oregon, Eugene, OR 97403, USA
             }

   \date{Received xxx; accepted xxx}
 
  \abstract
{The condensation of baryons within a dark matter (DM) halo during galaxy formation should result in some contraction of the halo as the combined system settles into equilibrium. We quantify this effect on the cuspy primordial halos predicted by DM-only simulations for the baryon distributions observed in the galaxies of the SPARC database. We find that the DM halos of high surface brightness galaxies (with $\Sigma_{\rm eff}\gtrsim100$ $L_\odot$ pc$^{-2}$ at 3.6 $\mu$m) experience strong contraction. Halos become more cuspy as a result of compression: the inner DM density slope increases with the baryonic surface mass density. We iteratively fit rotation curves to find the balance between initial halo parameters (constrained by abundance matching), compression, and stellar mass-to-light ratio. The resulting fits often require lower stellar masses than expected for stellar populations, particularly in galaxies with bulges: stellar mass must be reduced to make room for the DM it compresses. This trade off between dark and luminous mass is reminiscent of the cusp-core problem in dwarf galaxies, but occurs in more massive systems: the present-epoch DM halos cannot follow from cuspy primordial halos unless (1) the stellar mass-to-light ratios are systematically smaller than expected from standard stellar population synthesis models, and/or (2) there is a net outward mass redistribution from the initial cusp, even in massive galaxies widely considered to be immune from such effects.}

   \maketitle


\section{Introduction}
Dark matter (DM) halos are thought to play a central role in the formation and evolution of galaxies. During the early stage of galaxy formation, DM halos accrete gas into their gravitational potential wells, which is then converted into stars and eventually re-ejected in the circum-galactic medium by feedback processes. The initial halos are completely self-supported, so the halo structure can be determined through DM-only simulations. \citet{Navarro1996} simulated 19 DM halos ranging from dwarf galaxies to rich clusters. In spite of the big varieties in mass and size, the simulated halos, once scaled by characteristic volume density ($\rho_s$) and radius ($r_s$), can be described by a universal density profile, that is the Navarro-Frenk-White (NFW) model,
\begin{equation}
\frac{\rho}{\rho_s} = \frac{1}{\frac{r}{r_s}(1+\frac{r}{r_s})^2}.
\end{equation}
A key characteristic of the NFW profile is the presence of a central DM ``cusp'': for $r\rightarrow 0$, $\rho \rightarrow \infty$.

The NFW model has been extensively tested in galaxies using rotation curves and mass models. For high-mass and high-surface-brightness galaxies, it is often possible to obtain satisfactory fits to the rotation curves with NFW halos \citep[e.g.,][]{Katz2017} thanks to the degeneracy between the stellar mass-to-light ratio and the halo parameters \citep{vanAlbada1985}: the stellar contribution can be appropriately tuned to make adequate room for the inner DM cusp. For low-mass and low-surface-brightness, instead, the DM halo generally dominates the dynamics down to small radii, so the stellar contribution plays a minor role: the observed rotation curve shape is largely driven by halo profile. These faint galaxies generally show slowly rising rotation curves that contradict the NFW model, which predicts that rotation curves should rise steeply \citep[e.g.,][]{deBlok2001, deBlok2002, deBlok2008}. This contradiction is well-established as one of the big challenges of the cold DM model: the cusp-core problem \citep[e.g.][]{Moore1994,McGaugh2001,deNaray2009,Oh2011}.

As DM halos accrete baryons, there are two basic types of baryonic processes that may potentially alter the halo structure: adiabatic contraction and feedback. Adiabatic contraction is the response of DM halos to the gravitational potential of accreted baryons \citep{Sellwood2014}. Feedback is a term used to generically describe processes that return energy to the inter-steller medium (ISM), such as radiation pressure, stellar winds, and supernovae \citep{Agertz2013}. Adiabatic contraction concentrates the dark matter halos from its initial distribution while feedback may have the opposite effect. Opinions vary as to how effective feedback can be in redistributing dark matter \citep{Natarajan1999,MacLow1999,Efstathiou2000,Mo2004,Mashchenko2008,Keres2009,Dutton2009,Sawala2010,Governato2010,Scannapieco2012,Onorbe2015,Read2016,Read2016b,Katz2018,Bose2019}, and some even argue whether it can do this at all \citep{Gnedin2002,deNaray2011,Bland-Hawthorn2015,McGaugh2021b}. Nevertheless, there is widespread consensus that feedback acts in the right direction to help remedy the cusp-core problem provided that the energy injected by compacted sources is coupled to the ISM with high efficiency \citep{Bullock2017}.

A specific example of simulations in which feedback leads to a beneficial modification of DM halos is provided by \citet{DC2014, DC2014b}. In their simulations, feedback creates cores in initially cuspy dark matter halos over a broad range of masses. The resulting halo model (`DC14') provides good fits to rotation curve data \citep{DiCintio2016,Katz2017,Li2020}. This action has a sweet spot in the stellar-to-halo mass ratio around $\log (M_*/M_h) \approx -2.3$. This creates cores in the DM halos of dwarf galaxies with maximum effect around $M_* \approx 2 \times 10^8\;\mathrm{M}_{\sun}$. Core creation occurs over a broad mass range, but not over the entire mass range exhibited by galaxies. At very small masses, so little star formation has occurred that there is little feedback, thus a pristine NFW halo should persist \citep[but see also][]{Read2016}. At the large masses appropriate to Milky Way-like spiral galaxies, the gravitational potential becomes too large for feedback to have much impact, so one expects NFW-like DM halos. This seems to be a widespread expectation for massive galaxies. Their rotation curves, indeed, present satisfactory fits with the NFW model, but these fits generally do not consider the effect of adiabatic compression on the DM halo \citep{Barnes1984,Blumenthal1986,Jesseit2002,Gnedin2004,Sellwood2005}. Standard rotation-curve fits, indeed, treat DM halos and baryonic disks as fixed, independent components that set the gravitational potential, without considering their mutual gravitational interaction.

In this paper, we break the problem down into its component pieces, and choose to focus on the aspect for which we can make a rigorous computation: the contraction of the dark matter halo in response to the growth of the baryonic disk. Specifically, we numerically compute the adiabatic contraction of DM halos for the observed distribution of baryons in galaxies in the SPARC database \citep{SPARC}. Section 2 describes the algorithm that is used to consider the adiabatic contraction of DM halos and its application to the SPARC galaxy sample; Section 3 highlights the importance of adiabatic compression in rotation-curve fits; Section 4 introduces a new approach to fitting rotation curves that implements the adiabatic contraction of dark matter halos; Section 5 summarizes the results of this paper.

\section{Method and data}

\subsection{The algorithm for computing compression}

The contraction of DM halos in response to baryons can be modeled as an adiabatic process \citep{Choi2006}. Its study was pioneered by \citet{Blumenthal1986} \citep[also see][]{Barnes1984, Ryden1987}. Blumenthal's algorithm assumes that shells of matter do not cross, which implies particles in the same spherical shell contract or expand as a whole. Random motions of particles along radial direction are effectively ignored. This would not affect the evolving mass density profile as long as particles moving in and out each shell are well balanced, but this proves to be an oversimplification. Blumenthal's method conserves only angular momentum, but adiabatic process should conserve all three actions of an orbit \citep{Binney1987}. Several authors \citep{Barnes1987, Sellwood1999, Gnedin2004} showed that Blumenthal's algorithm results in more compression than their N-body simulations show. The assumption that shells do not cross ignores radial motions that mitigate the amount of compression.

\citet{Young1980} introduced a method that shows better consistency with N-body simulations by conserving all three adiabatic actions,
\begin{equation}
    J'_r(E', L) = J_r(E, L);\ \ J_\phi=L;\ \ J_\theta=0,
    \label{AdiaAct}
\end{equation} 
where $J_r$ and $J'_r$ are the radial actions before and after compression, respectively; $E$ is the total energy (kinetic energy + gravitational potential energy) of dark matter particles; $L$ is the angular momentum perpendicular to the baryonic disk; $J_\theta$ is the azimuthal adiabatic action. The main difference between Young's and Blumenthal's methods is the conservation of radial action $J_r$. \citet{Young1980} showed that isotropic distribution is less compressed than purely circular motions, so radial motions effectively make halo compression more difficult. The adiabatic action in azimuth is simply the angular momentum, which is conserved in both methods. The $\theta$ component is essentially zero, as the disk plane of the considered galaxy is assumed invariant during adiabatic contractions. Young's method assumes spherical symmetry throughout the contraction process, which significantly simplifies the algorithm. \citet{Jesseit2002} and \citet{Sellwood2005} tested this assumption with numerical simulations, and found that the computed density profile is consistent with the spherical average of simulated N-body halos.

Young's method starts with a pure DM system with given volume density profile $\rho(r)$, potential function $\Phi(r)$, and particle distribution function $f(E, L)$. We adopted for these quantities a pristine, primordial NFW halo. Baryons were then added using the observed surface mass density profile by gradually increasing the normalization factor. This way, we built a final distribution that matches the azimuthal average of a real galaxy \citep[see][]{Sellwood2005}. The gravitational potential changes as
\begin{equation}
    \Phi_{\rm tot}(r) = \Phi_{\rm DM}(r) + \Phi_{\rm Bar}(r),
\end{equation}
where $\Phi_{\rm Bar}(r)$ increases in small increments. The size of each increment is chosen to make sure the change of gravitational potential is small enough for a perturbative method. This helps achieve convergence and allows the use of the adiabatic approximation in which the adiabatic actions remain invariant, so one can derive the updated total energy $E'$ according to equation \ref{AdiaAct}. The distribution function can then be updated based on the conservation in the number of particles:
\begin{equation}
    f'(E', L) = f(E, L).
\end{equation}
With the updated potential function and distribution function, one can update the density profile of the DM halo according to
\begin{equation}
    \rho'(r) = 4\pi\int^{\Phi(\infty)}_{\Phi(r)}\int^{L_{\rm max}}_{0}\frac{Lf'(E', L)}{r^2v_r}dLdE,
\end{equation}
where $v_r$ is the radial velocity. The new density profile can then be used to recalculate the potential function, and the whole process is iterated until two successful potential functions are consistent at a satisfactory level.

\begin{figure*}
    \centering
    \includegraphics[scale=0.4]{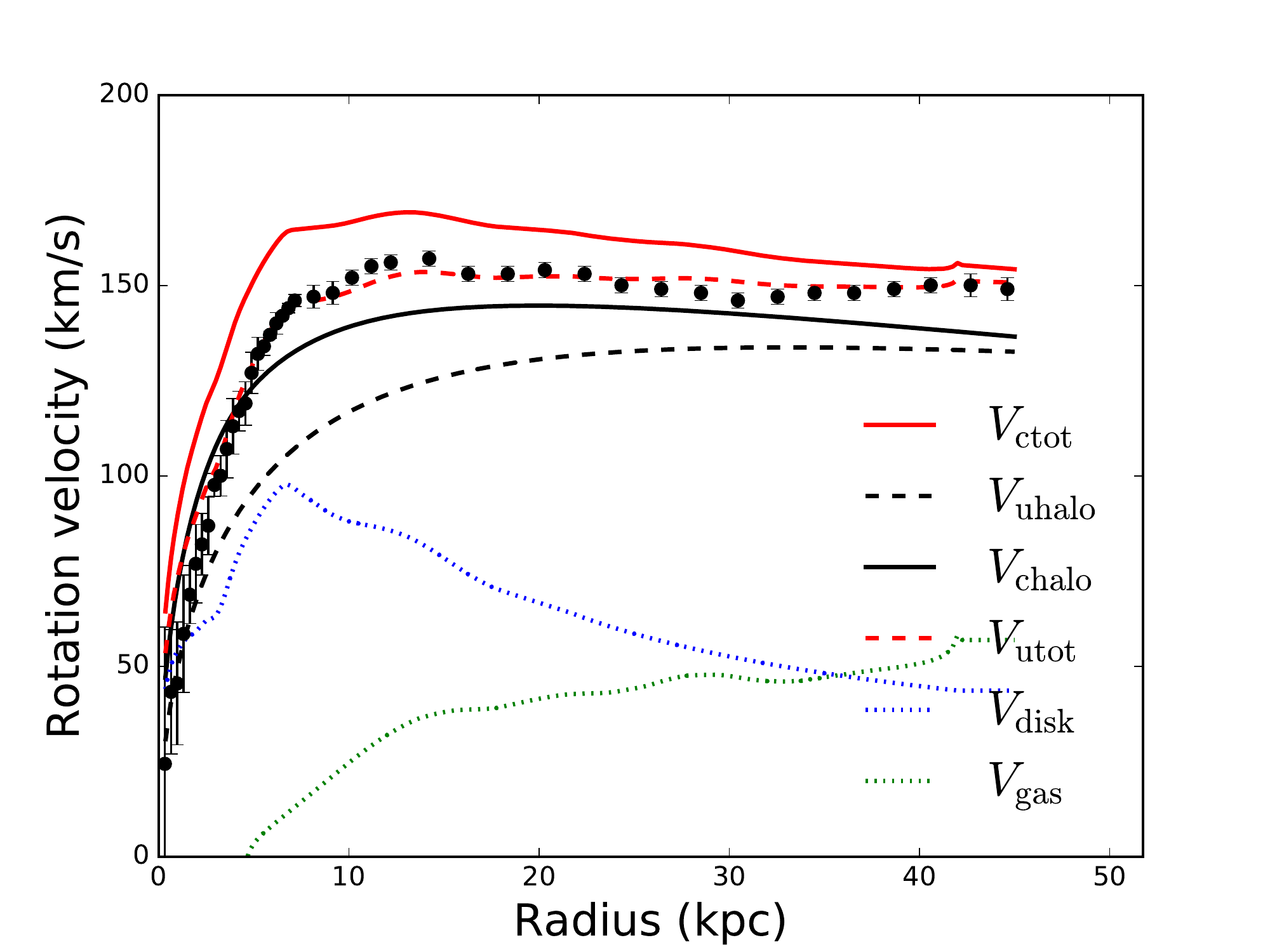}\includegraphics[scale=0.4]{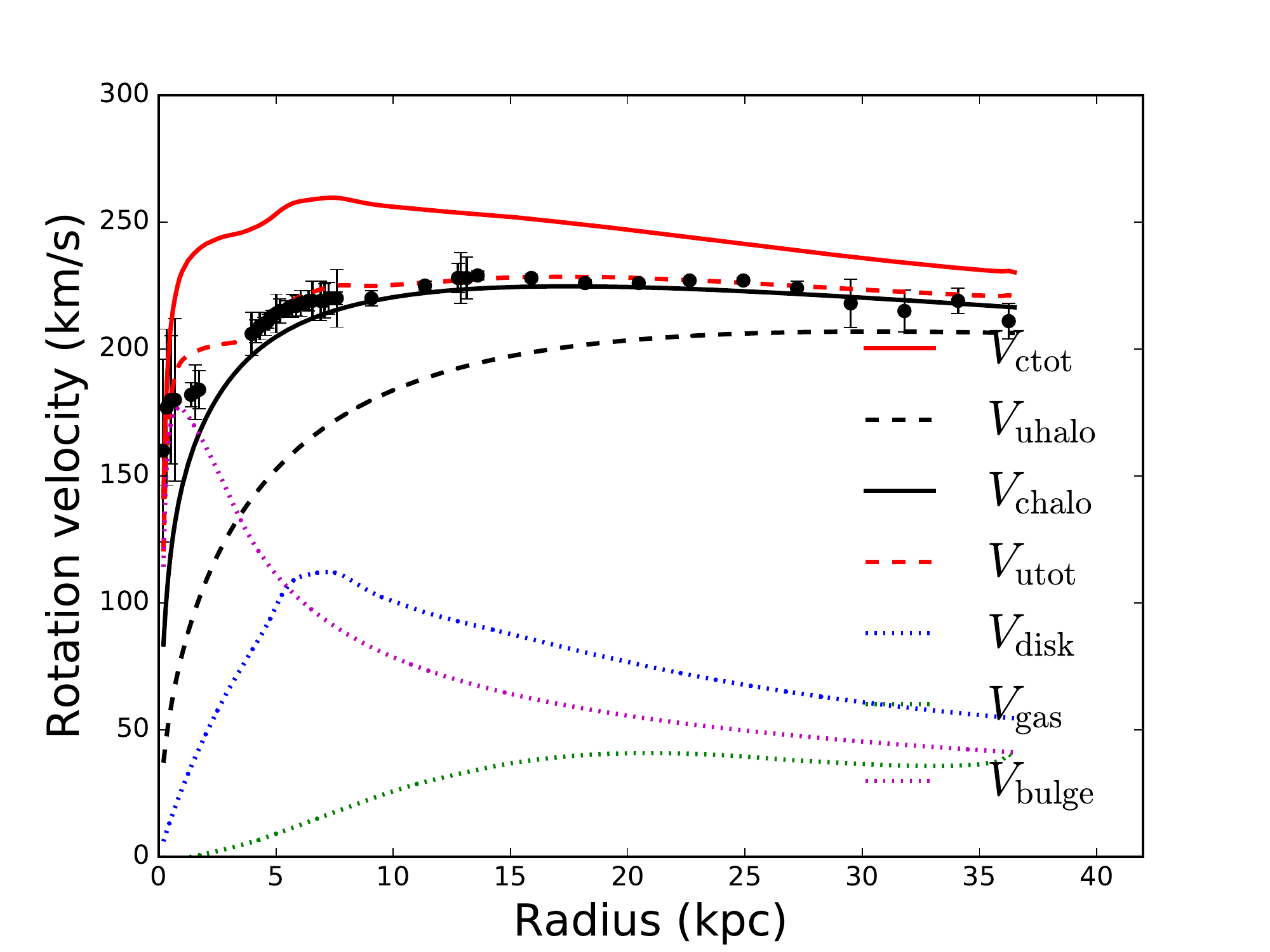}
    \caption{Examples of compressed DM halos in NGC 3198 (left) and UGC 06786 (right). Points with error bars are the observed rotation curves, adjusted to the best-fit disk inclinations. Green, blue, and purple dotted lines represent the contributions of gas, disk, and bulge, respectively, which have been adjusted according to the best-fit stellar mass-to-light ratios and galaxy distances as in \citet{Li2020}. Black dashed lines show the contribution of best-fit ``static'' NFW halos, while red dashed lines are the total rotation curve fits from \citet{Li2020}. Black solid lines show the contributions of compressed DM halos, and red solid lines are the corresponding total rotation curves. Compressed halos overshoot the data in both cases. The magnitude of the compression is illustrated by the difference between the dashed and solid black lines. Halo compression is a non-negligible effect in massive galaxies: the realistic dark matter halo must differ from a purely NFW form.}
    \label{fig:compression}
\end{figure*}

One of the advantages of Young's algorithm for adiabatic compression as implemented by \citet{Sellwood2005} is the ability to work directly with observational data, such as the surface brightness for disk, bulge and gas at different radii. With a choice of stellar mass-to-light ratio, this specifies a baryonic mass distribution that matches a real galaxy. For the dark matter halo, one needs to input two characteristic parameters to specify the initial NFW halo. After reading the initial halo and baryonic mass distribution, the code iterates the potential function until its variation is smaller than $10^{-3}$, providing as output the compressed halo profile custom-made by the specified baryon distribution. There is no guarantee that the initial characteristic parameters for the halo will be a match to the observed rotation curve, but this process is computationally efficient so it is possible to iterate to obtain a fit (\S \ref{sec:fitting}). In this way, it is in principle possible to infer the parameters of the dark matter halo that is the primordial antecedent to the observed, post-compression halo. This helps bridge the chasm between the predictions of dark matter-only simulations and observational reality.

\subsection{Data: The SPARC sample}

This project requires multiple types of data. We need to know (i) the total gravitational potential and (ii) the baryonic mass distribution for each and every galaxy in a sample that fairly represent the entire parameter space over which galaxies exist in morphology, mass, size, etc. The gravitational potential can be traced by rotation curves obtained with optical spectroscopy or radio interferometry, while the baryonic mass includes both stars and gas which must be traced by separate observations in different parts of the spectrum (typically optical or near-infrared images for the stars and 21 cm line data for the gas). All of these various data must be obtained for all of the galaxies in the sample. No ideal sample exists, but the Spitzer Photometry \& Accurate Rotation Curves \citep[SPARC,][]{SPARC} database comes as close as has so far been achieved \citep{McGaugh2020IAUS}. In this paper, we apply Sellwood's compression code to a large number of SPARC galaxies.

The SPARC database combines extended H {\footnotesize I} rotation curves from interferometric observations obtained by many independent workers \citep[see][]{SPARC}. The rotation curves are very extended, providing a tracer of the gravitational potential from the centers of galaxies to large radii. These galaxies have [3.6] images in the \emph{Spitzer} archive that are used to derive their surface brightness profiles. This provides an excellent map of the stellar mass \citep{Schombert2019}. In this work, we construct mass models sampled with the full resolution of the native photometry to map the baryonic gravitational potentials of the galaxies\footnote{We exclude six galaxies (D512-2, D564-8, D631-7, NGC4138, NGC5907, UGC06818) from 175 that lack H{\footnotesize I} surface density profiles.}.

The SPARC sample is quite representative, in the sense that it spans a wide range in surface brightness ($\sim$ 3 dex), stellar mass ($\sim$ 5 dex), and gas fraction (from a few percent to over 90\%). This samples the range of properties exhibited by galaxies better than many larger samples, which are biased towards $L^*$ galaxies \citep{McGaugh2020IAUS}. We have explored the SPARC sample in a series of previous papers. In \citet{Li2020} we made rotation curve fits to the 175 galaxies in the SPARC database using seven halo models, and determine the properties of their DM halos. The NFW fits from this work provide the starting point for our current discussion.

\section{Illustrating the magnitude of the compression: Motivations to incorporate baryonic contraction when fitting rotation curves}\label{sec:compression}

In the traditional approach to fitting rotation curves, one chooses a dark matter halo profile (e.g., NFW) and finds the parameters that best match the data in combination with the observed stars and gas. This provides an analytic approximation to the distribution of dark matter over the observed radial range of the data. However, the dark matter halo must respond to baryonic dissipation by adjusting its distribution. It is uncertain how significant the response is. As a first step, we computed the magnitude of adiabatic compression with the best-fit DM halos of \citet{Li2020}. At this juncture, we were not attempting to fit the rotation curve data (see \S \ref{sec:fitting}), and simply wished to illustrate the effects of compression. This provided a direct hint on if it is necessary to consider the mutual gravitational interaction between dark matter and baryons when fitting rotation curves.

\begin{figure*}
    \centering
    \includegraphics[scale=0.4]{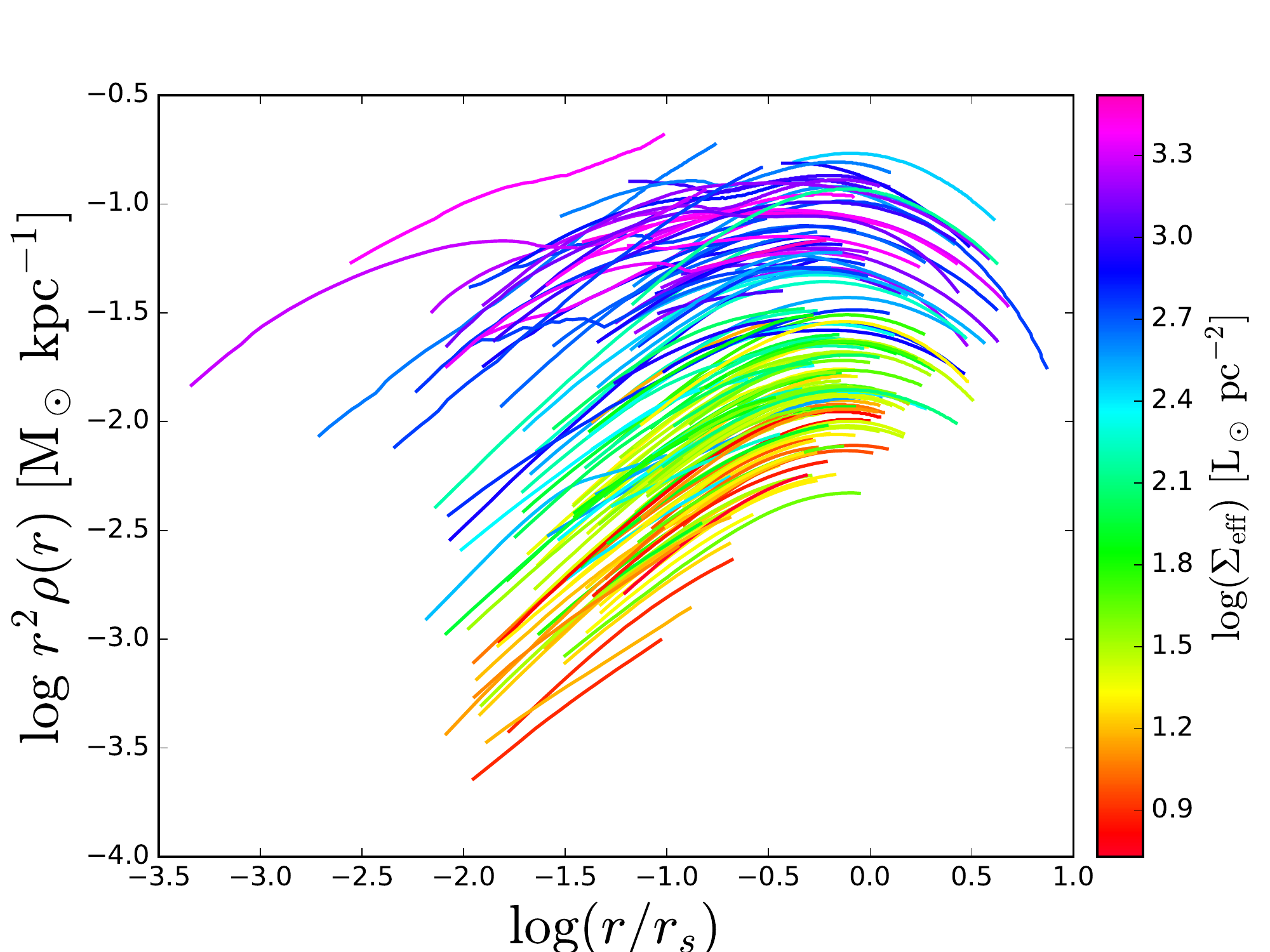}\includegraphics[scale=0.4]{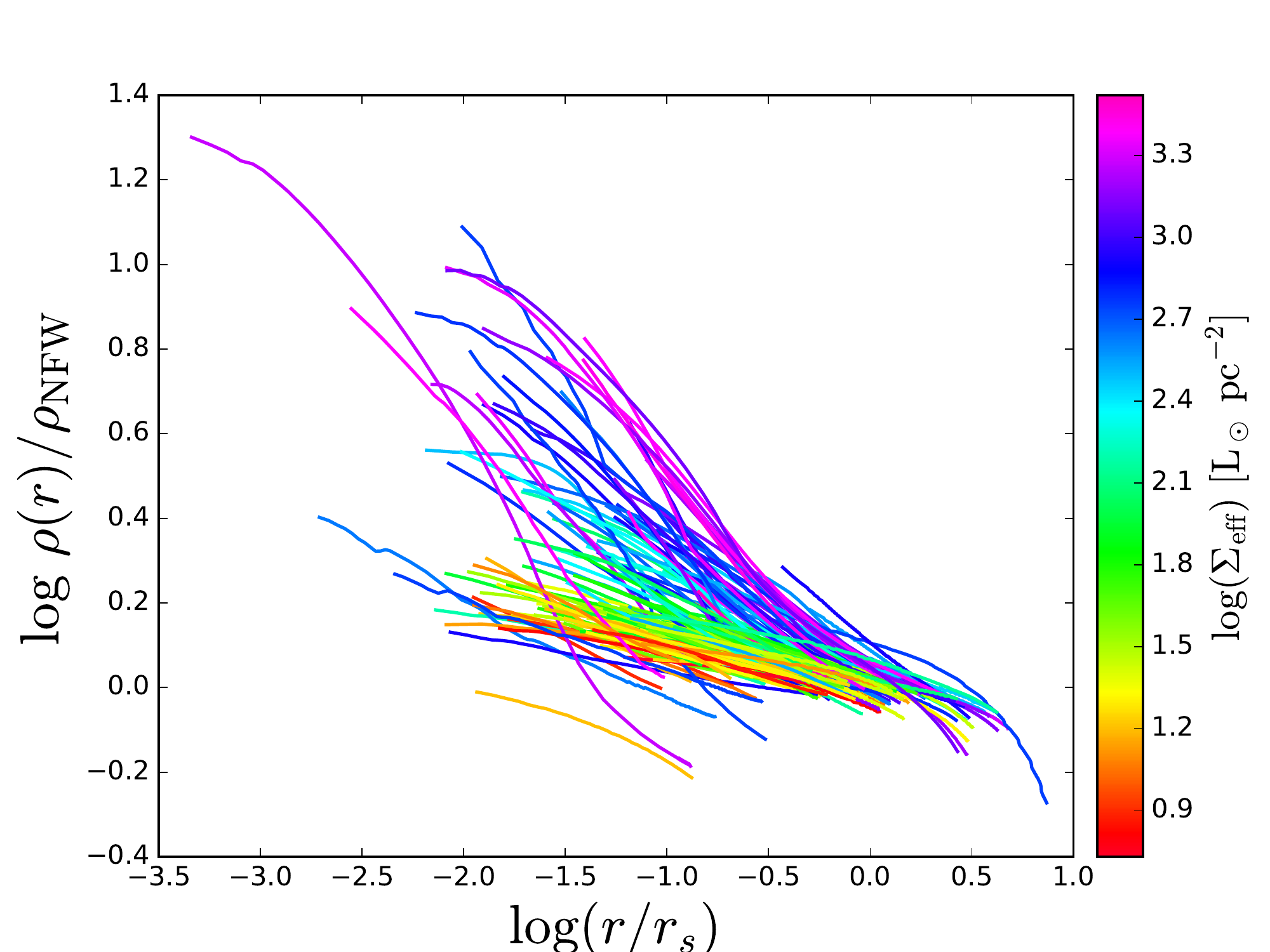}
    \caption{Baryonic compression on halo density profiles. Left: density profiles of compressed halos for the SPARC galaxies starting from the static NFW best-fit models. Right: density ratios of compressed to initial NFW halos. Both are color coded by effective surface brightness. High surface-brightness galaxies generally show larger density ratios, indicating more adiabatic contraction. Density ratios also increase towards the centers of galaxies where bulges can have a pronounced effect.}
    \label{fig:denratio}
\end{figure*}

\citet{Li2020} fit rotation curves of SPARC galaxies with a wide variety of dark matter halo models using a Bayesian approach with either flat priors or $\Lambda$CDM priors from the halo mass--concentration relation and the stellar mass--halo mass relation. For our current investigation, we started with the fits obtained for NFW halos with \LCDM\ priors. In addition to fitting the halo parameters, \citet{Li2020} allowed the mass-to-light ratio to vary with a stellar population prior, and treated the distance and inclination of each galaxy as nuisance parameters. Here we adopt these best-fit parameters, adjusting each parameter accordingly from their nominal values in the raw SPARC database. These adjustments are usually small, as the uncertainties in these observed quantities were imposed as a prior.

Starting from the best-fit NFW halos, we used COMPRESS \citep{Sellwood2005,Sellwood2014} to derive the evolution of DM halos in response to the incremental addition of baryons. The resulting output is a DM halo that is in dynamically equilibrium with the observed baryonic distribution, but may in general not fit the observed rotation curve anymore.

Figure \ref{fig:compression} presents as examples the rotation curves of two high-surface-brightness galaxies, NGC 3198 and UGC 06786. The fits of \citet{Li2020} for these galaxies are satisfactory, given the steeply rising rotation curves of high-surface-brightness galaxies can accommodate the NFW model. However, computing the compression without reassessing the initial conditions make the total contributions overshoot the data (Fig.\ \ref{fig:compression}). The difference is more substantial at small radii, where the effect of the baryons is maximized. Consequently, the DM halo is no longer NFW in form.

Figure \ref{fig:denratio} plots the density profiles of the compressed halos for the entire SPARC sample, color-coded by the effective surface brightness of each galaxy. We find a rainbow distribution of density profiles, as higher-surface-brightness galaxies have higher DM halo densities. In order to illustrate the differences between the initial and compressed halos, we plot their density ratios in the right panel of Figure \ref{fig:denratio}. We find a similar rainbow distribution in terms of surface brightness. This implies that higher surface-brightness galaxies have more substantial compression, as expected: a greater concentration of baryons leads to a greater compression of the DM halo.

\begin{figure*}
    \centering
    \includegraphics[scale=0.4]{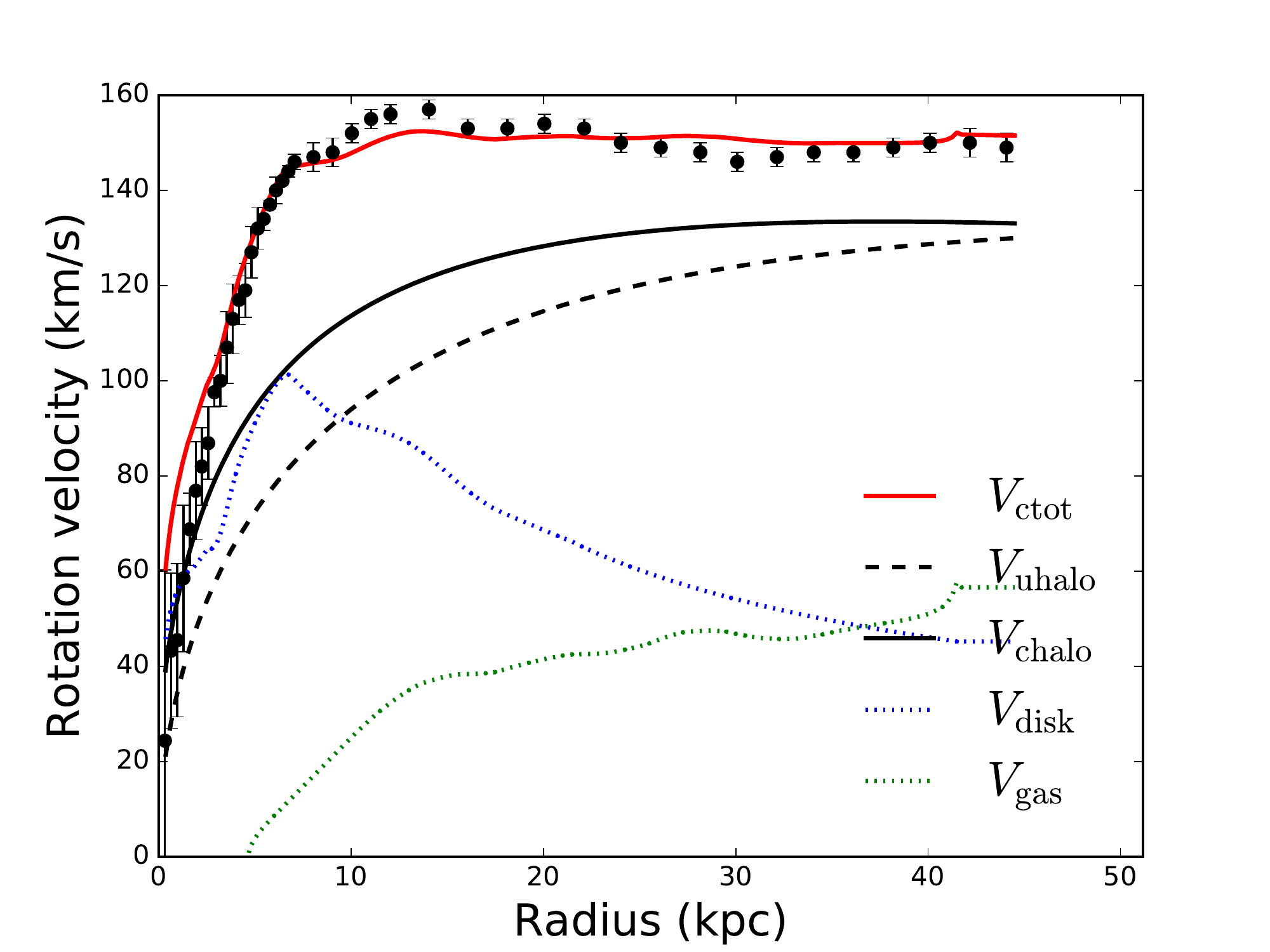}\includegraphics[scale=0.4]{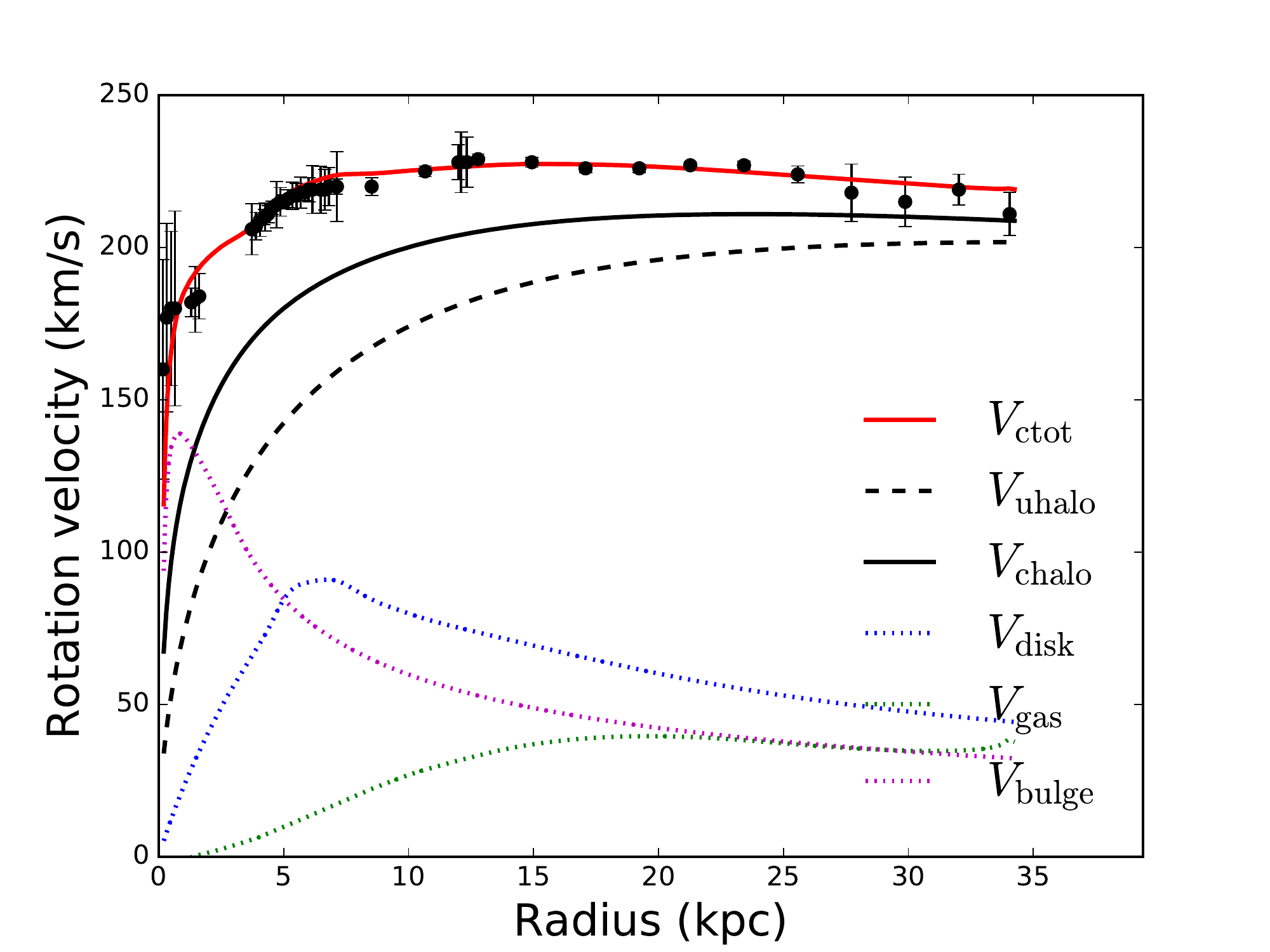}\\
      \includegraphics[scale=0.4]{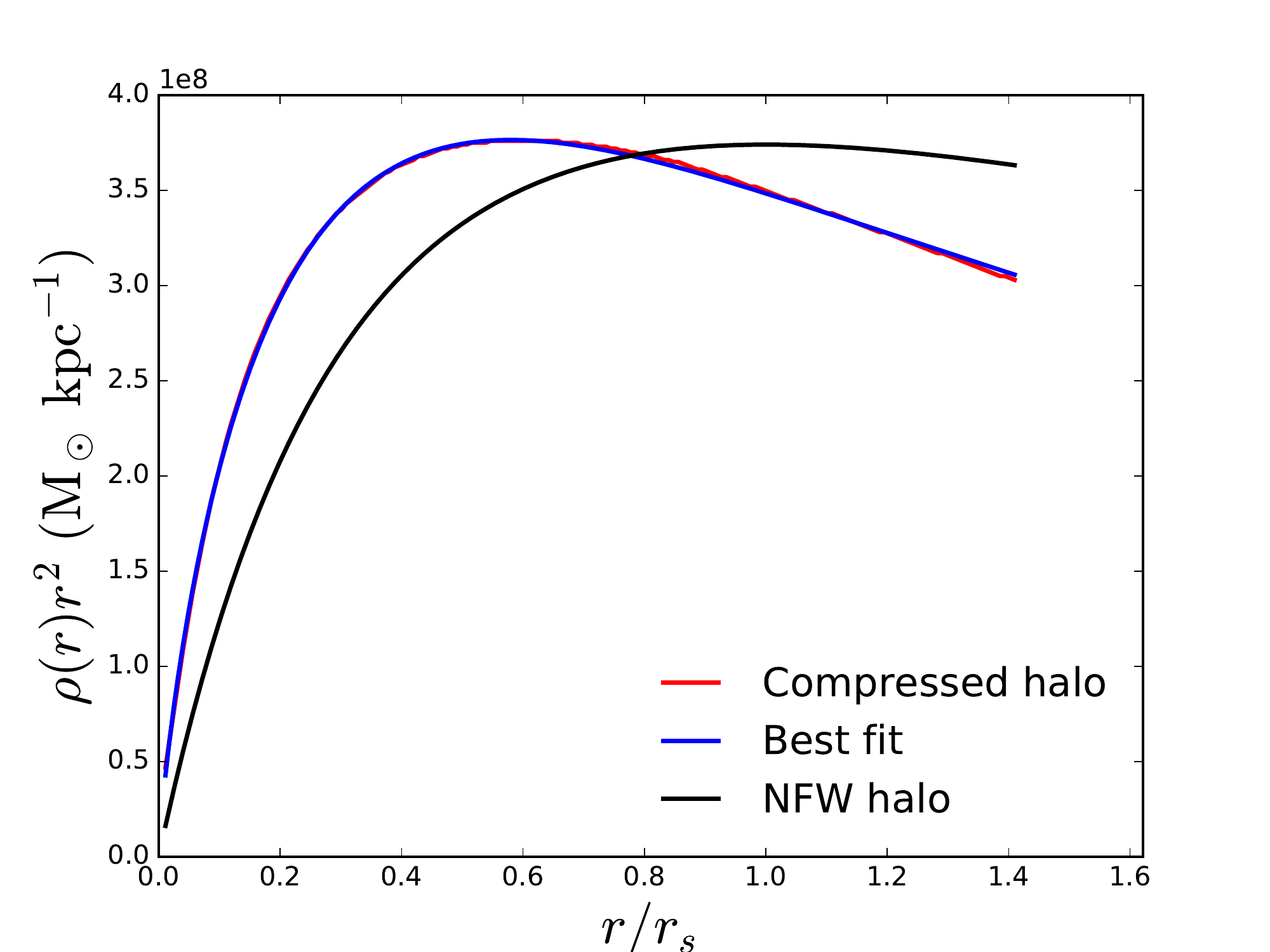}\includegraphics[scale=0.4]{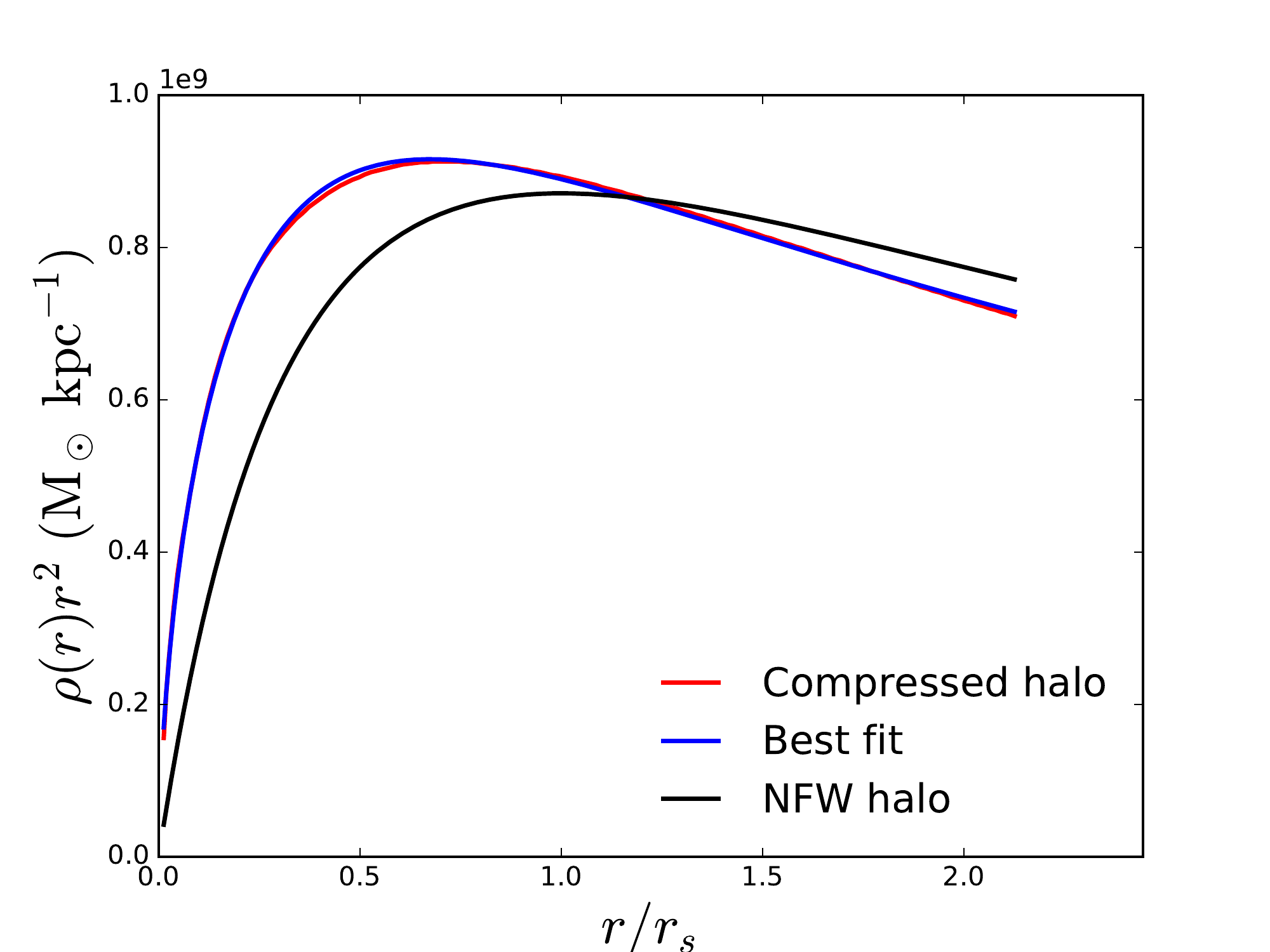}
    \caption{Example rotation curve fits and halo density profiles. Top: rotation-curve fits with adiabatically compressed DM halos for NGC 3198 (left) and UGC 06786 (right). Green, blue, and purple dotted lines represent the contributions of gas, disk and bulge, respectively. Black solid lines show the best-fit compressed DM halos, while black dashed lines show the corresponding initial NFW halos. Red solid lines show the expected total rotation curves, which match the data better than Figure \ref{fig:compression} after adjusting the fitting parameters. Bottom: The density profiles of the compressed (red lines) DM halos and associated initial NFW halos (black lines). Blue lines show a parametric fit to the compressed DM halo using the ($\alpha$, $\beta$, $\gamma$) model with $\alpha$ fixed to unity. The best-fit parameters for stellar disks, bulges, prior-compression and post compression DM halos are presented in Table \ref{tab:parameters}. The complete figure set of 125 rotation curve fits and density plots are available in the SPARC database.}
    \label{fig:fit}
\end{figure*}

Indeed, the compression effect is strongly surface brightness dependent. It is modest for low-surface-brightness galaxies, where the increase in the dark matter density is minor. This makes sense, as there is little baryonic mass for the dark matter to respond to in these galaxies, which are nearly all straight lines in the right panel of Fig.\ \ref{fig:denratio}. As the surface brightness increases up to around $\Sigma_{\rm eff}=100$ $L_\odot$ pc$^{-2}$, there is a transition in the shape of these curves to include a sharp upturn in the ratio of final to initial dark matter density at small radii. This always happens in galaxies with bulges, but also occurs in some bulgeless galaxies with high surface brightness disks. The common feature seems to be maximality \citep{Starkman2018}: the compression effect becomes very strong when $V_{\rm Bar} \rightarrow V_{\rm obs}$. This again makes sense, as it is natural that the compression due to baryons should be most pronounced where they dominate the gravitational potential.

The strong response of the halo in regions of baryon domination immediately highlights a problem for fits of halos with cusps like the NFW model. Existing fits of this type are generally obtained for a static halo by reducing the stellar mass-to-light ratio from the expectation of stellar population synthesis in order to accommodate the cusp \citep{Li2020}. This tension seems mild until we realize that halos are not static entities. In reality, the dark matter must respond to the growth of the baryonic gravitational potential, and the response is strong where the baryonic concentration is greatest. As we have seen, ignoring this effect would result in best-fit NFW halos that are not in dynamical equilibrium with embedded baryons. One has to fit rotation curves and compress DM halos simultaneously.

\section{Fitting rotation curves with compressed halos}\label{sec:fitting}

\subsection{Fitting strategy}

Since the standard approach to fitting rotation curves does not account for adiabatic compression, we devised a new approach that treats DM halos and baryonic disks as a coupled system. Specifically, we combined adiabatic contraction with rotation curve fitting by iterating the compression code with different initial inputs. For a given stellar mass-to-light ratio, we started with some assumed parameters for the initial NFW halos, and computed their stabilized density profiles through adiabatic contraction. We then compared the compressed halos with observed rotation curves. The comparison provides a feedback that is used to determine the parameters of the initial NFW halos for next iteration.

This procedure is much more computationally expensive than standard rotation-curve fits with fixed DM halos. To reduce the parameter space to be explored and gain computational speed, we kept galaxy distance and disk inclination fixed to the fiducial SPARC values and fit for stellar mass-to-light ratio $\Upsilon_\star$ and two halo parameters, $V_{200}$ and $C_{200}$, which are defined as
\begin{equation}
    C_{200} = r_{200}/r_s \;;\; V_{200} = 10\ C_{200}r_s H_0,
\end{equation}
where $r_{200}$ is the virial radius that encloses a mean halo density equal to
200 times the critical density of the universe.

\begin{figure*}
    \centering
    \includegraphics[scale=0.4]{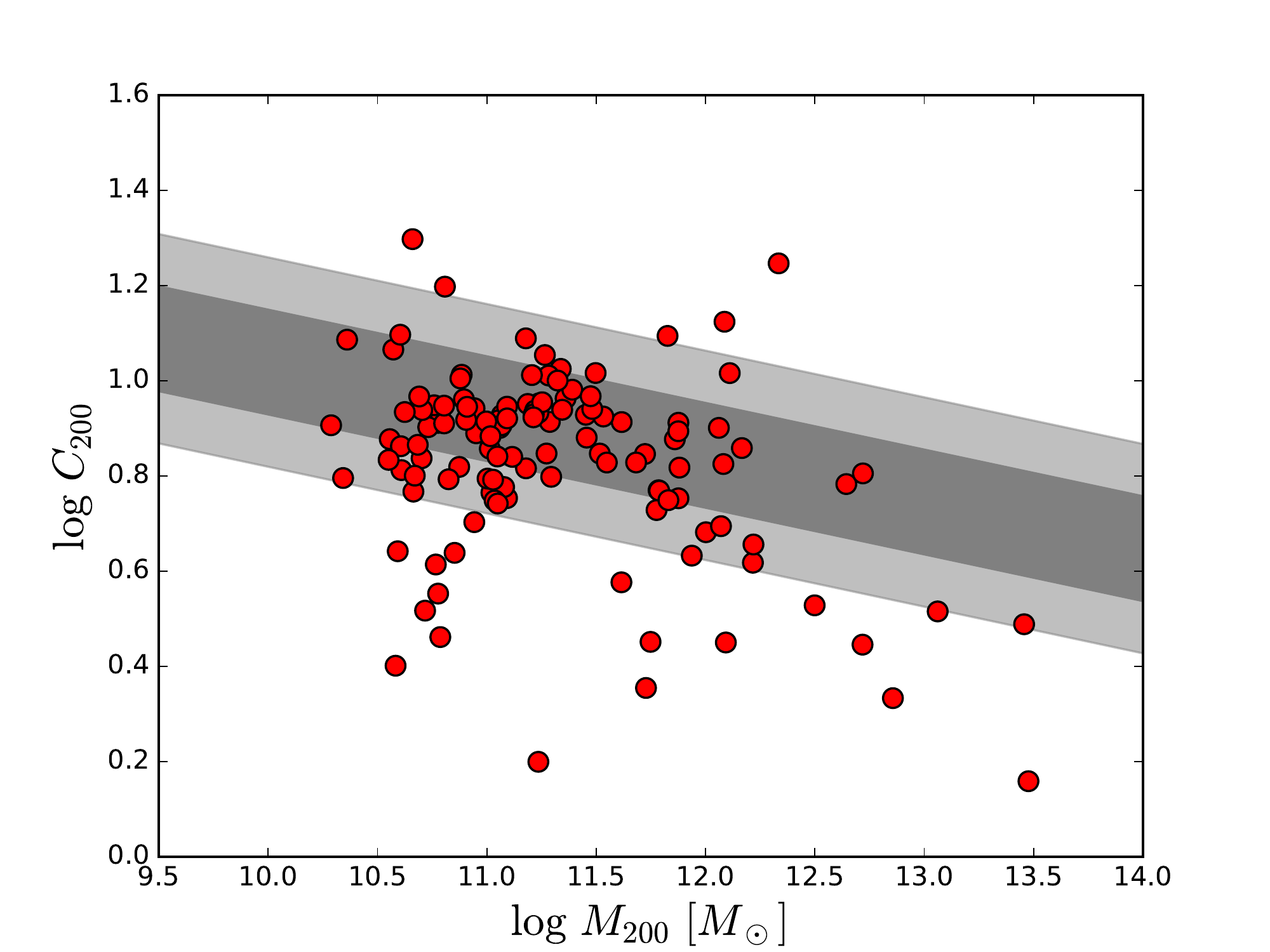}\includegraphics[scale=0.4]{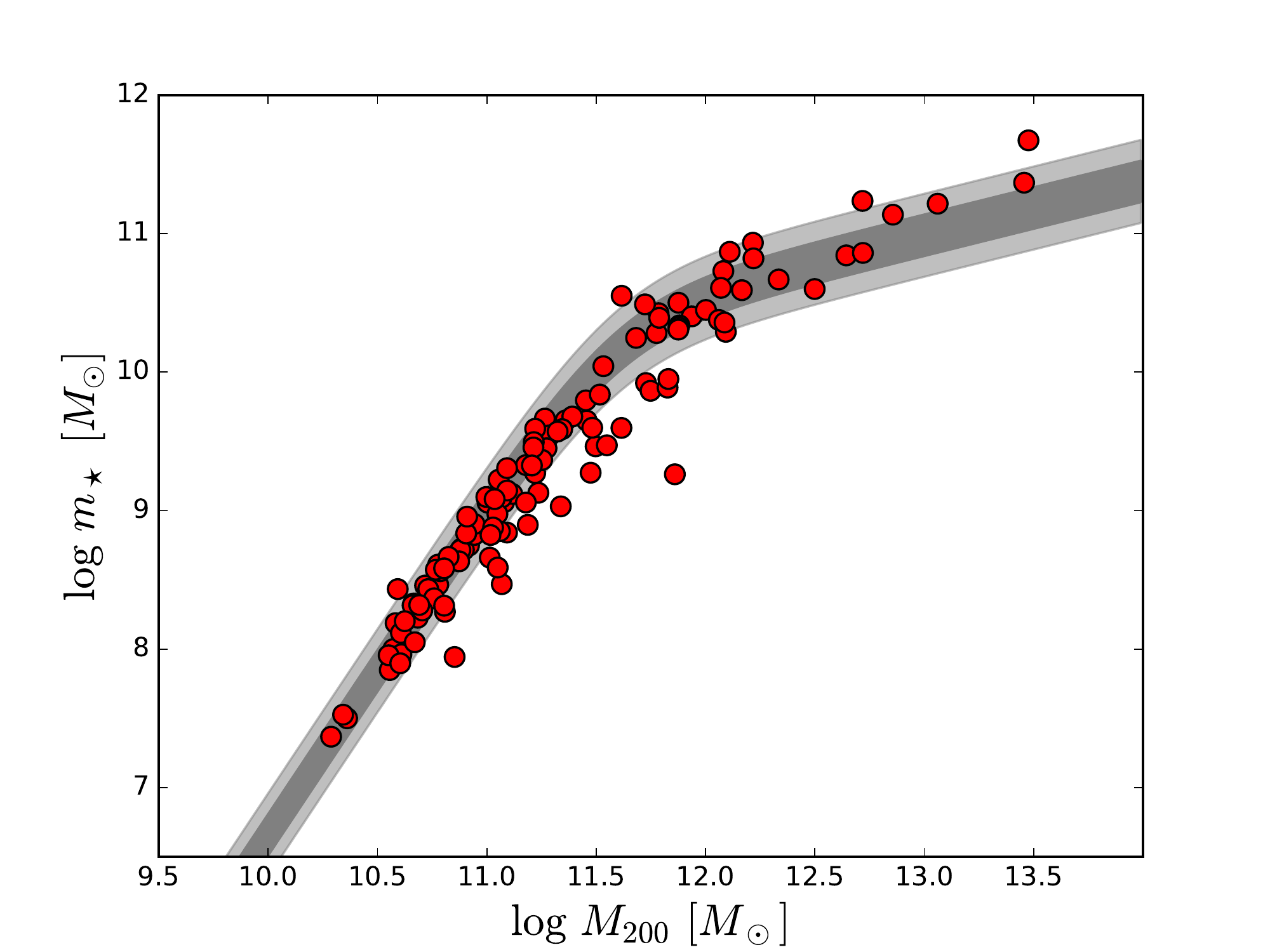}
    \caption{\LCDM\ priors: the halo mass-concentration relation (left panel) from DM-only simulations \citep{DuttonMaccio2014} and the stellar-to-halo mass relation (right panel) from abundance matching \citep{Moster2013}. In both panels, red circles show the best-fit halo masses and concentrations of the initial NFW halos. Dark and light gray bands show the expected 1$\sigma$ and 2$\sigma$ regions, respectively. Both relations are imposed as priors in fitting rotation curves. Our results show larger scatters than the simulation predictions.}
    \label{fig:LCDMprior}
\end{figure*}

\begin{figure*}
    \centering
    \includegraphics[scale=0.4]{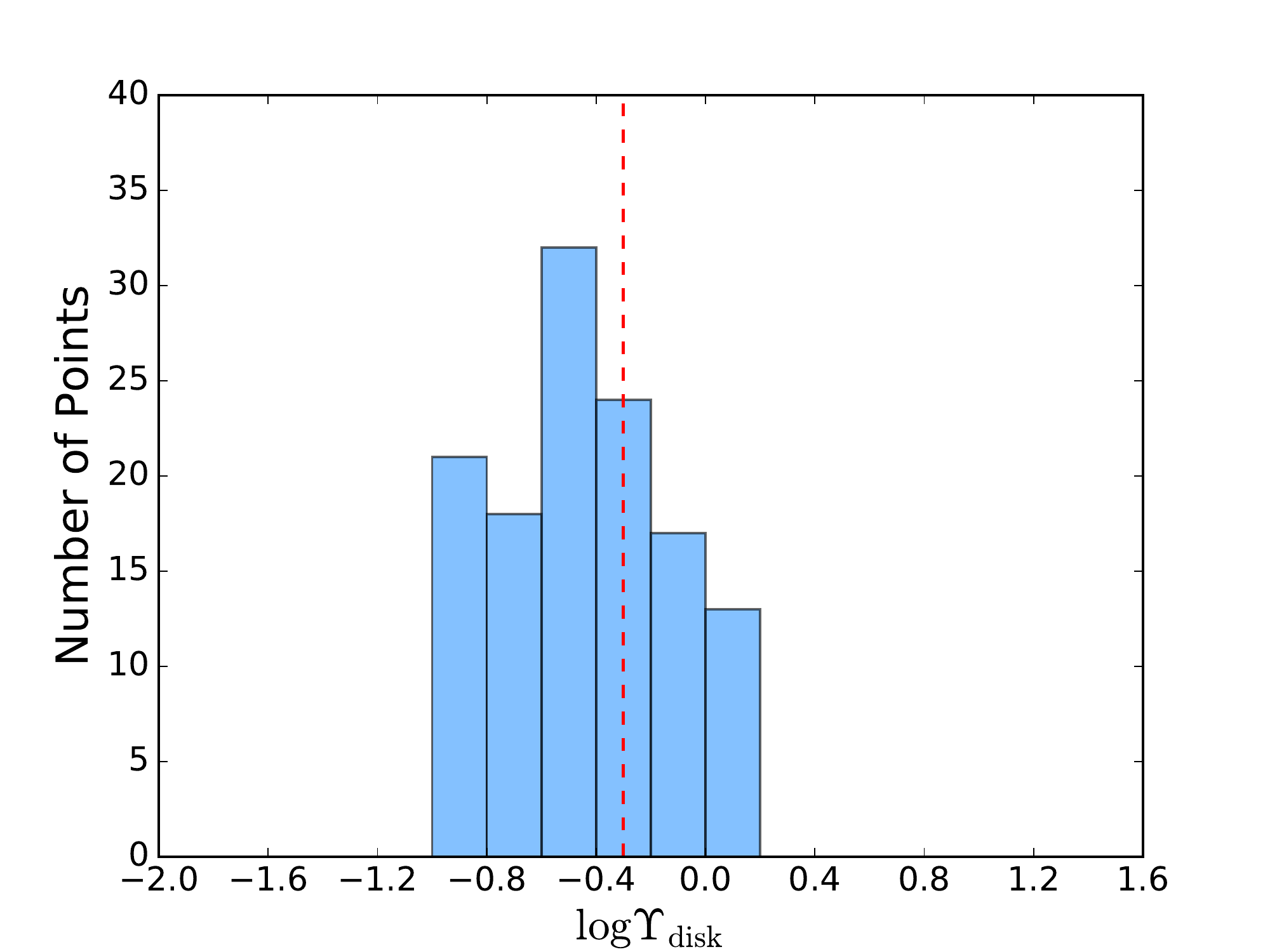}\includegraphics[scale=0.4]{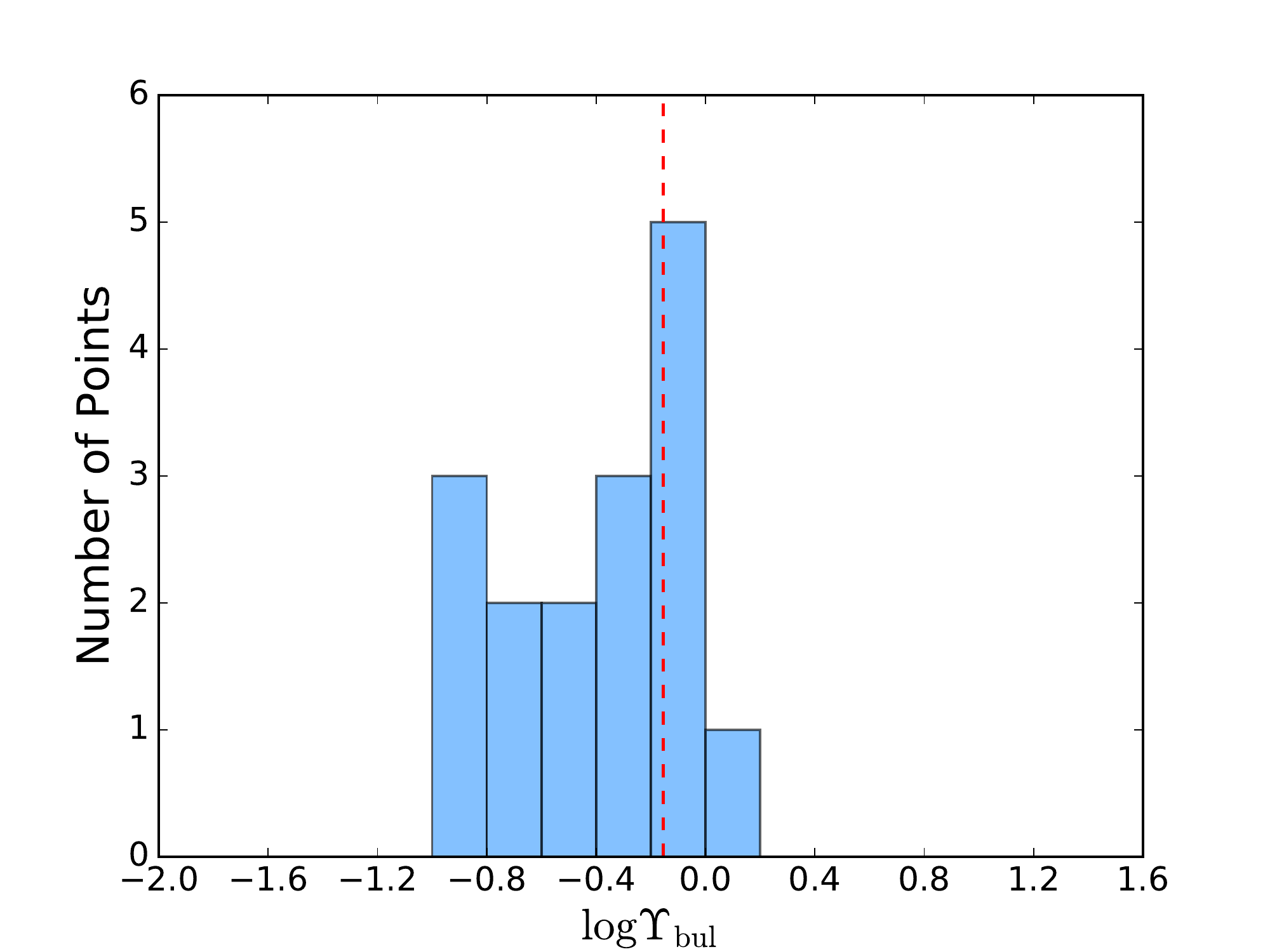}
    \caption{Histograms of best-fit stellar mass-to-light ratios for disks and bulges. Vertical red dashed lines indicate the fiducial values $\Upsilon_{\rm disk}=0.5$ and $\Upsilon_{\rm bulge}=0.7$ from stellar population synthesis models \citep{Schombert2019, Schombert2022}. The imposed hard boundaries are (0.1, 1.0). The best-fit stellar mass-to-light ratios spread over the whole range, but show a preference for smaller values than expected.}
    \label{fig:Ydisk_hist}
\end{figure*}

For each set of \{$V_{200}$, $C_{200}$, $\Upsilon_\star$\} values, we ran the compression code and obtain the corresponding compressed halo. The code outputed the halo contributions $V_{\rm DM}$ at each radii. The total rotation velocities are the summation of the contributions of DM halos and baryonic distributions,
\begin{equation}
    V^2_{\rm tot} = V^2_{\rm DM} + \Upsilon_{\rm disk}V^2_{\rm disk} + \Upsilon_{\rm bulge}V^2_{\rm bulge} + |V_{\rm gas}|V_{\rm gas},
\end{equation}
where $\Upsilon_{\rm disk}$ is the stellar mass-to-light ratio of the stellar disk and $\Upsilon_{\rm bulge}$ that of the bulge (if there). We then calculated the probability function,
\begin{equation}
    P=-\frac{1}{2}\chi^2 + {\rm prior}, 
\end{equation}
where $\chi^2$ quantifies the fit quality and prior considers the constraints on fitting parameters. The probability function serves as a feedback for each iteration. This helps determine a better set of fitting parameters for the next iteration, until we find the best halo parameters that maximize the probability function with the given stellar mass-to-light ratio $\Upsilon_{\star}$. We then choose a different value of $\Upsilon_{\star}$, and repeat the process. Eventually, we find the values of $\Upsilon_{\star}$, $C_{200}$ and $V_{200}$ corresponding to the maximum probability. The stellar mass-to-light ratio is varied within (0.1, 1.0) with an accuracy of 0.01. This is a wider range than the model predicts \citep[e.g.][]{BelldeJong2001}, but has larger flexibility to achieve satisfactory fits.

In this work, we imposed the $\Lambda$CDM prior, which is comprised of the abundance matching relation \citep{Moster2013} and the halo mass concentration relation \citep{Maccio2008, DuttonMaccio2014}. Both relations evolve with cosmic time. This is because DM halos grow themselves regardless of baryons. In reality, the growth of DM halos and the contraction should happen simultaneously. For simplicity, we treated these two processes separately, assuming the growth has been completed prior to baryonic compression. Since we were studying nearby galaxies, we imposed $\Lambda$CDM priors at $z=0$ on the NFW halos before compression: the stellar-halo mass relation 
\begin{equation}
    \frac{m_\star}{M_{200}} = 2N\Big[\Big(\frac{M_{200}}{M_1}\Big)^{-\beta} +  \Big(\frac{M_{200}}{M_1}\Big)^{\gamma}\Big]^{-1},
\end{equation}
where $\log M_1 = 11.590$, $N = 0.0351$, $\beta = 1.376$, $\gamma = 0.608$, and the halo mass-concentration relation \citep{Maccio2008},
\begin{equation}
    \log C_{200} = 0.830 - 0.098\log(M_{200}/[10^{12}h^{-1}M_\odot]),
\end{equation}
where $M_{200}$ is the enclosed halo mass defined at virial radius $r_{200}$. The abundance matching relation and the halo mass-concentration relation have an estimated scatter of 0.15 dex and 0.11 dex around the mean relations, respectively, which are used as the standard deviations in the lognormal priors.

\begin{figure*}
    \centering
    \includegraphics[scale=0.4]{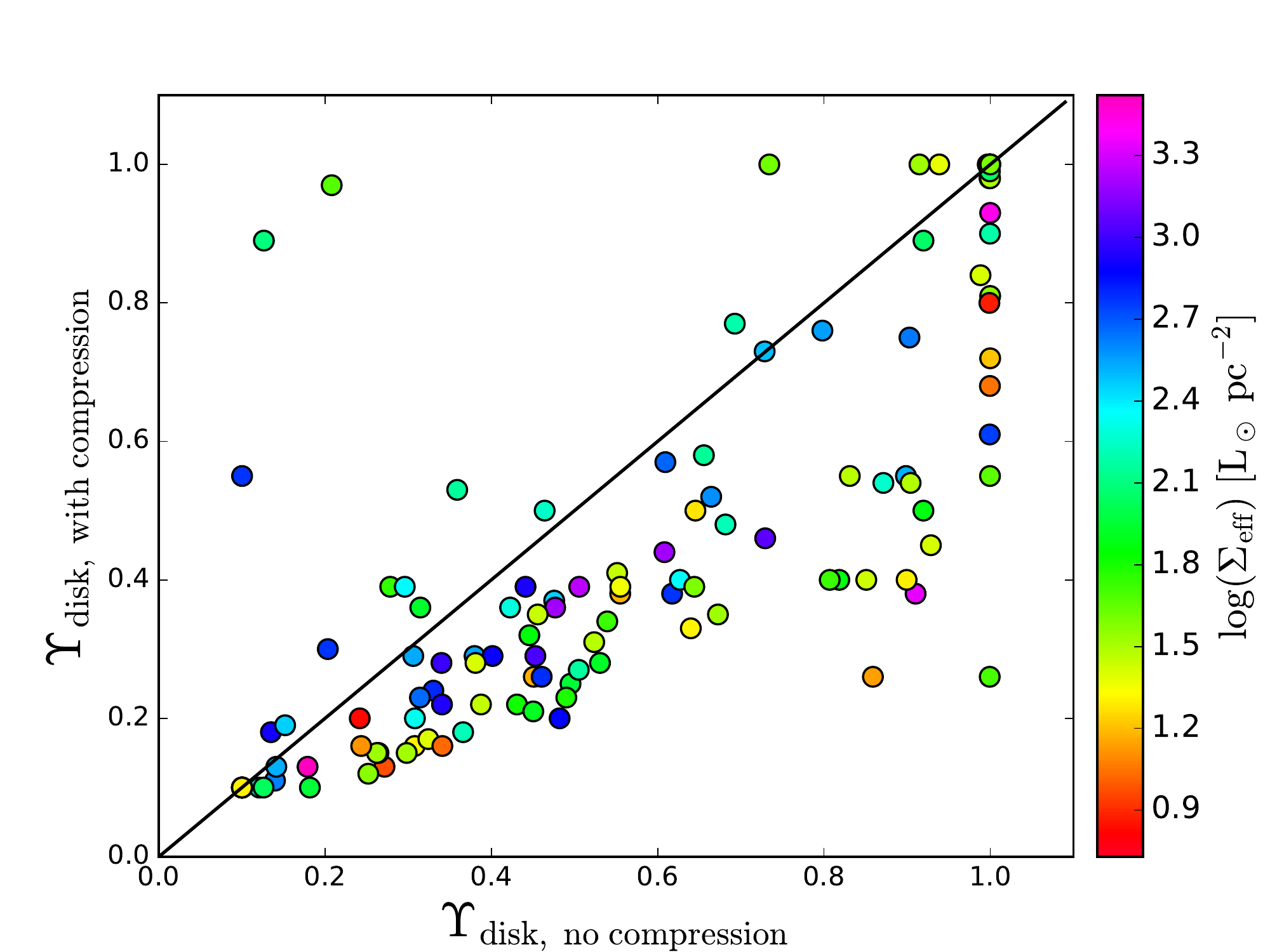}\includegraphics[scale=0.4]{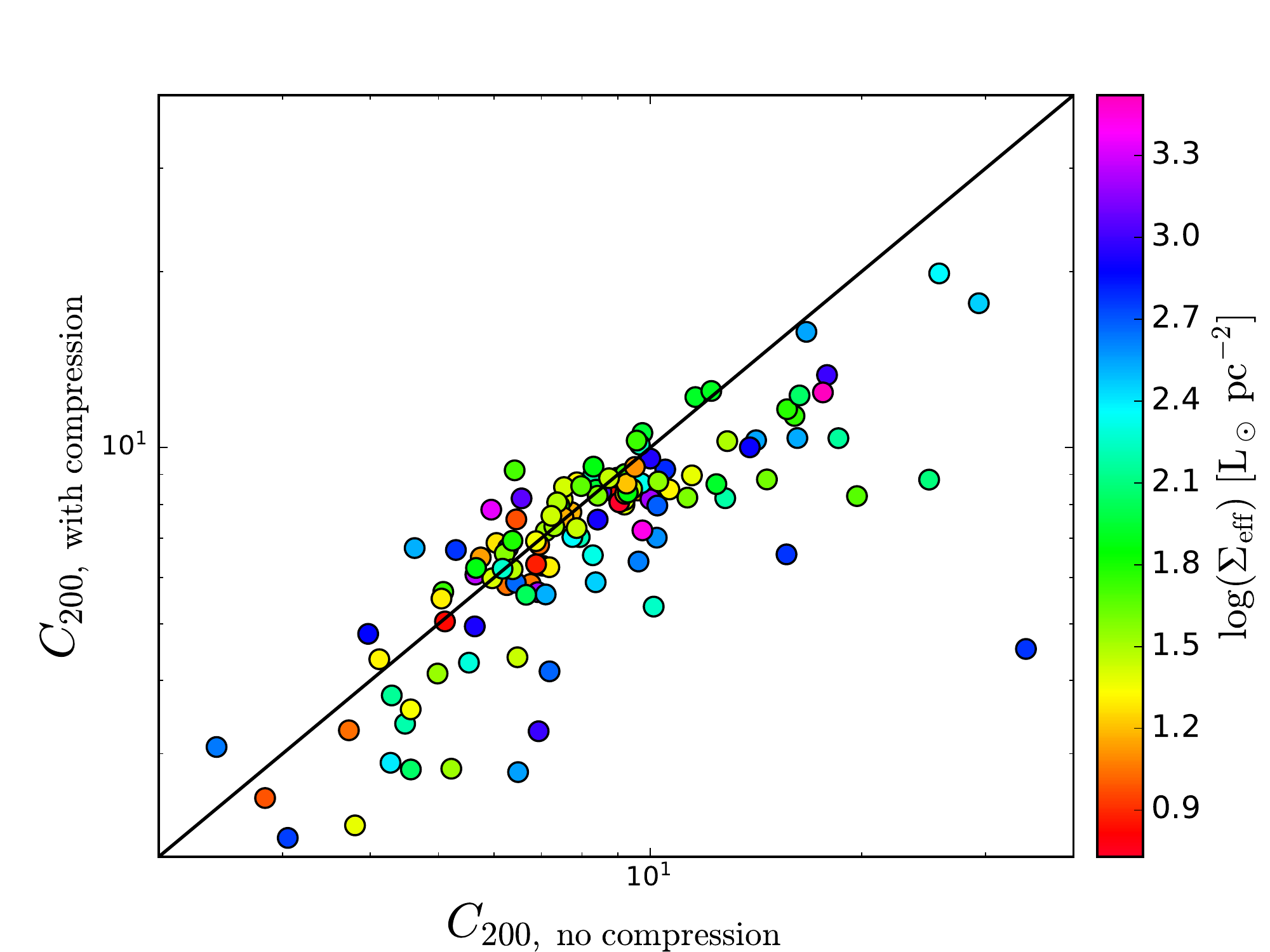}
    \caption{Effects of baryonic compression on fitting parameters. The comparison of the best-fit stellar mass-to-light ratios (left) and halo concentration (right) with and without adiabatic compression. Galaxies are color coded by their effective surface brightness. When baryonic contraction is taken into account, both the stellar mass-to-light ratios and the halo concentrations are reduced to leave room for more concentrated halos.}
    \label{fig:parameters}
\end{figure*}

\begin{figure}
    \centering
    \includegraphics[scale=0.4]{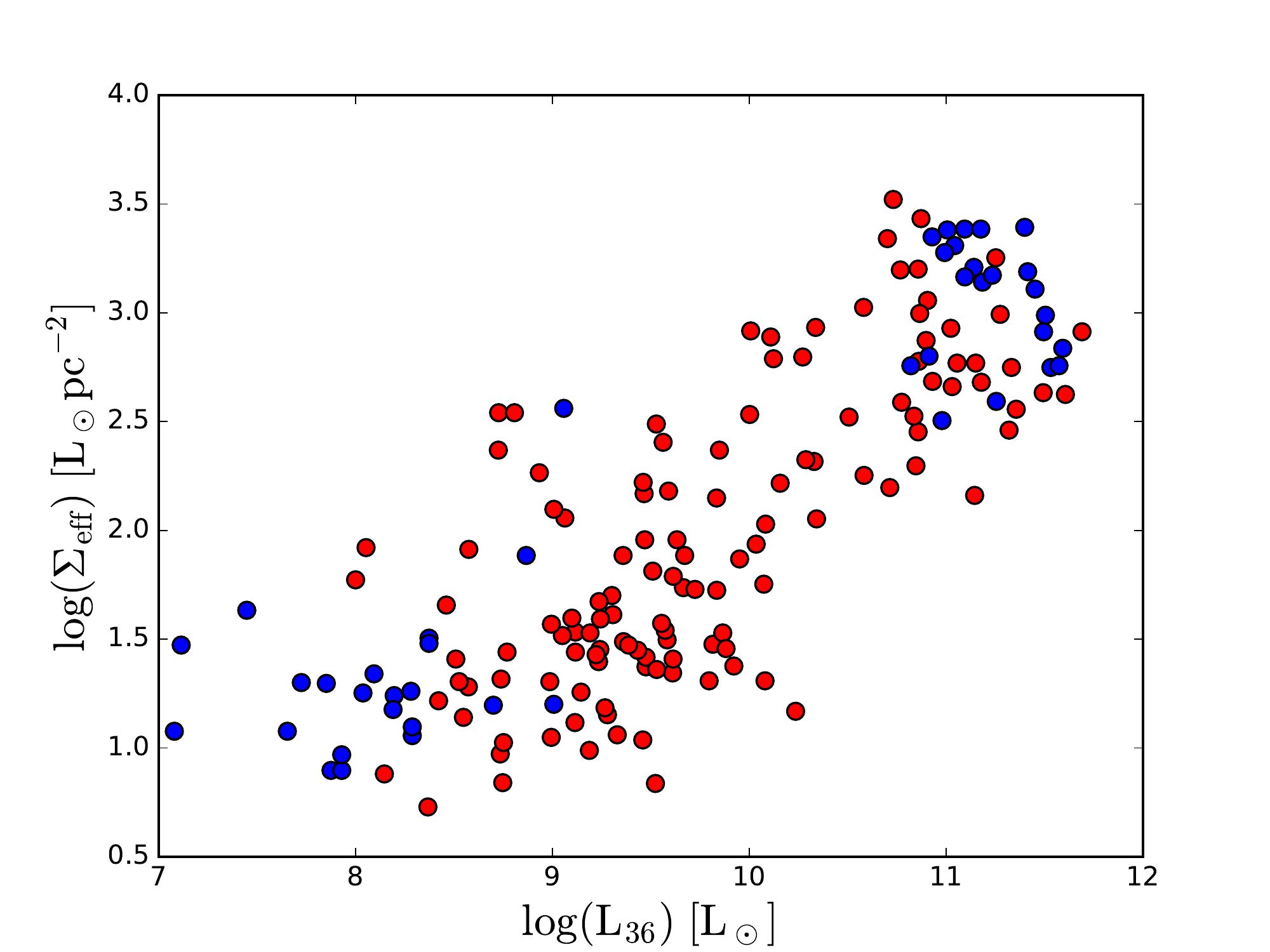}
    \caption{Parameter space of luminosity and effective surface brightness for the SPARC sample. Blue points mark the galaxies that do not have a converged DM halo in the presence of the baryonic gravitational potential, so that they cannot be fit by compressed halos.}
    \label{fig:GalaxyFail}
\end{figure}

\subsection{Examples of individual rotation-curve fits}

The top panels of Figure \ref{fig:fit} show two example fits. For better comparison, we choose the same galaxies as we did in Section \ref{sec:compression}. In both galaxies, the initial NFW halos are modified significantly: the final compressed DM halos provide much higher gravitational contributions at small radii. To obtain a satisfactory fit to the rotation curve, the halo concentration of NGC 3198 is reduced with respect to that in Figure \ref{fig:compression}, as demonstrated by the slowly flattening DM curve. For UGC 06786, both the bulge and disk contributions are decreased to leave room for the larger contribution of the compressed DM halo by adjusting their stellar mass-to-light ratios to lower values.

The bottom panels plot their prior-compression (namely NFW) and post-compression halo density profiles. The compressed halos present much higher densities at small radii with respect to the initial ones. This leads to larger contributions in the corresponding rotation curves. The compressed halos can be nicely fit with the general ($\alpha$, $\beta$, $\gamma$) model \citep{Hernquist1990, Zhao1996},
\begin{equation}
    \frac{\rho}{\rho_s} = \frac{1}{(\frac{r}{r_s})^{\gamma}[1+(\frac{r}{r_s})^{\alpha}]^{(\beta-\gamma)/\alpha}},
\end{equation}
where the transition parameter $\alpha=1$ is fixed. The fits are shown as blue lines in the bottom panels, which nicely capture both the inner and outer slopes. 

\subsection{Overview of the best-fit parameters}

We now check whether the fitting parameters \{$M_{200}$, $C_{200}$, $\Upsilon_\star$\} compare to the imposed \LCDM\ priors as well as the expectations of stellar population synthesis models. We stress that $M_{200}$ and $C_{200}$ refer to the initial NFW halos, not the compressed halos. In general, halo mass does not change during the baryon-driven contraction, but the halo concentration does. We considered the properties of initial NFW halos before compression because there are well-defined predictions from DM-only simulations.

Figure \ref{fig:LCDMprior} shows the halo mass-concentration relation and the stellar mass-halo mass relation. Both relations are broadly recovered, but the data present larger scatter than expected. A significant fraction of initial NFW halos have systematically smaller halo concentrations and larger halo masses than expected.

Figure \ref{fig:Ydisk_hist} shows the distribution of the best-fit stellar mass-to-light ratios. The best-fit values show a wide distribution that spans all the allowed range. We do not observe a lognormal distribution centered around $\Upsilon_{\rm disk} \simeq 0.5$ and $\Upsilon_{\rm disk} \simeq 0.7-0.8$ as expected from stellar population synthesis models \citep[e.g.][]{BelldeJong2001, Schombert2019, Schombert2022}. Instead, many galaxies prefer smaller values of $\Upsilon_{\rm disk}$ and $\Upsilon_{\rm bulge}$. This is likely due to the fact that the fitting routine must tune down the mass-to-light ratio in order to (1) make room for the cuspy NFW halo that is then compressed into an even cuspier halo, (2) minimize the baryon-driven compression as much as possible to reproduce the observed rotation curve.

\subsection{Comparison with the traditional approach}

In order to demonstrate the effect of adiabatic contraction, we repeated the fitting process using a Markov Chain Monte Carlo method \citep{Foreman-Mackey2013} but removed the implementation of adiabatic contraction. So we fit the same parameters and imposed the same priors. The results hence provide a baseline that can be used to investigate how baryonic compression could affect the resultant parameters.

We find that considering or neglecting baryonic compression does not have considerable effect on fit qualities, but leads to systematic changes in the distributions of fitting parameters. Figure \ref{fig:parameters} plots the best-fit stellar mass-to-light ratios and halo concentrations with and without baryonic compression. We do not show the comparison of $V_{200}$ because it is less sensitive to the local structure. Figure \ref{fig:parameters} shows systematically lower stellar mass-to-light ratios and halo concentrations when the baryonic compression is taken into account. This is expected since compressed halos generally make larger contributions to rotation curves than the original NFW halos, so that it requires either the baryonic contribution to be smaller or the initial NFW halos to be less concentrated. Varying stellar mass-to-light ratio has a more significant impact: 1) a smaller stellar mass-to-light ratio directly scales down the stellar contributions to the total rotation curves; 2) it also reduces the total baryonic mass and thereby makes halo contraction less profound. As such, reducing stellar mass-to-light ratios can help fit rotation curves quite effectively. Reducing the concentrations of the initial NFW halos provides further room for adiabatic contraction. Eventually, one can obtain comparably satisfactory rotation curve fits as if there was no adiabatic contraction. 

In spite of the similar fit qualities, the resultant fitting parameters differ significantly. This further suggests that the adiabatic contraction has to be properly taken into account when fitting rotation curves in order to solidly test DM models.

\begin{figure*}
    \centering
    \includegraphics[scale=0.4]{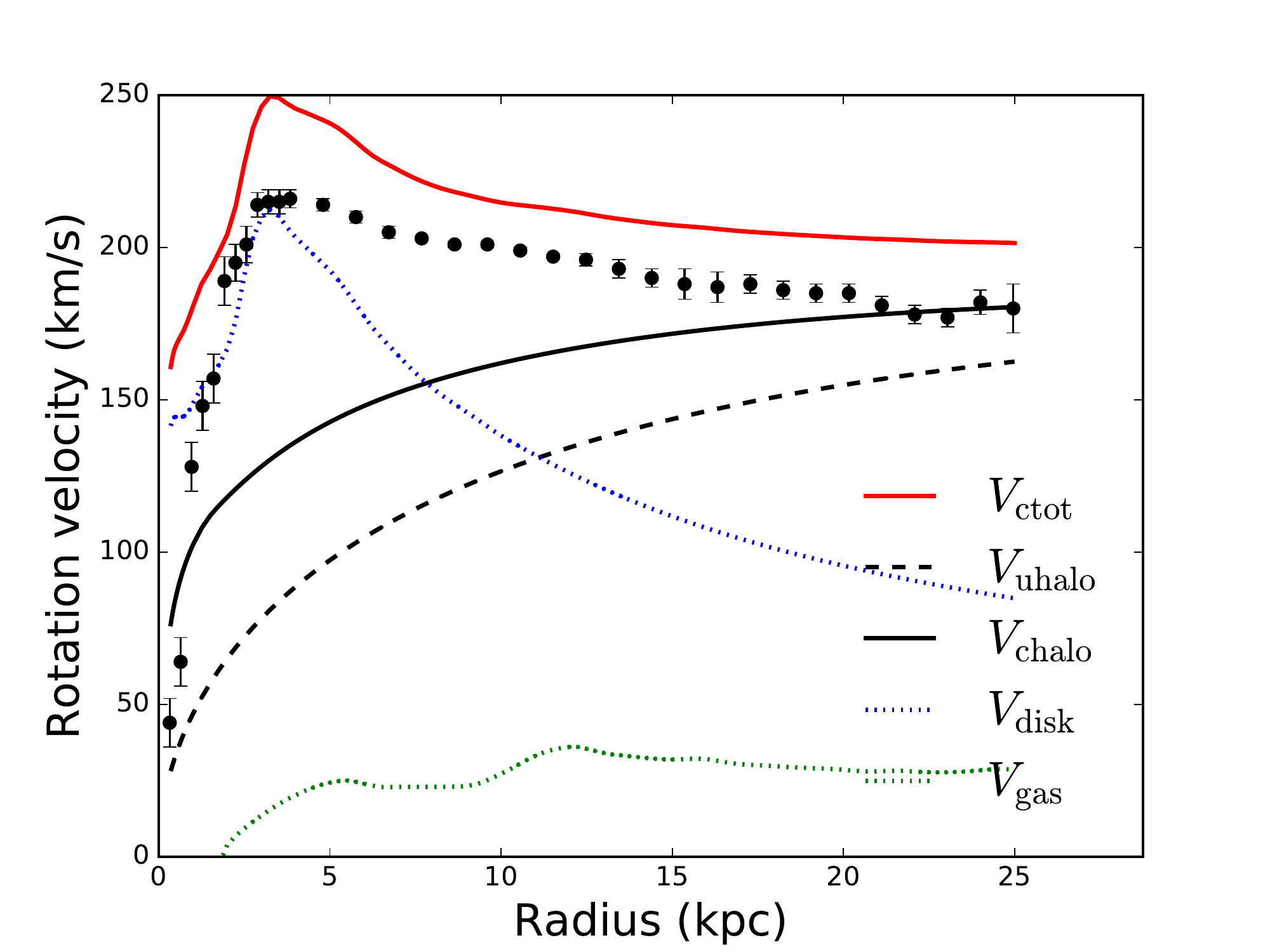}\includegraphics[scale=0.4]{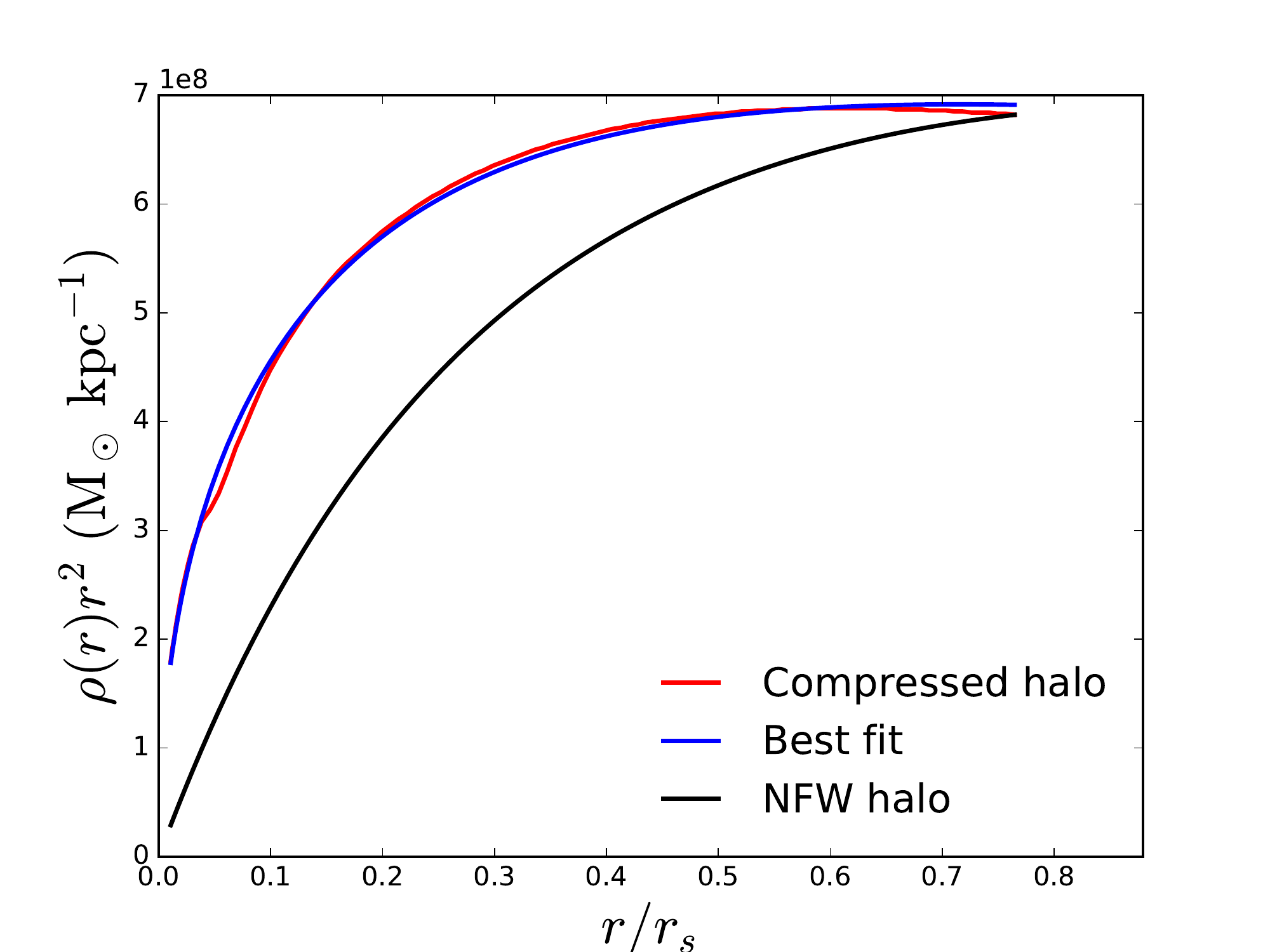}
    \caption{Compression on prior halos for failed galaxies. The rotation curve (left) of NGC 2903 and its DM halo density profiles (right). The green and blue lines are the gas and stellar disk contributions, respectively. The black dashed line is the $\Lambda$CDM predicted NFW halo assuming the abundance matching by \citet{Moster2013} and the halo mass-concentration relation by \citet{DuttonMaccio2014}. The black solid line is the corresponding compressed halo. This example shows how a \LCDM prior halo fit the data after baryonic effect is taken into account. The compressed halo shows significantly higher DM density at small radii.}
    \label{fig:FailExample}
\end{figure*}

\begin{figure*}
    \centering
    \includegraphics[scale=0.4]{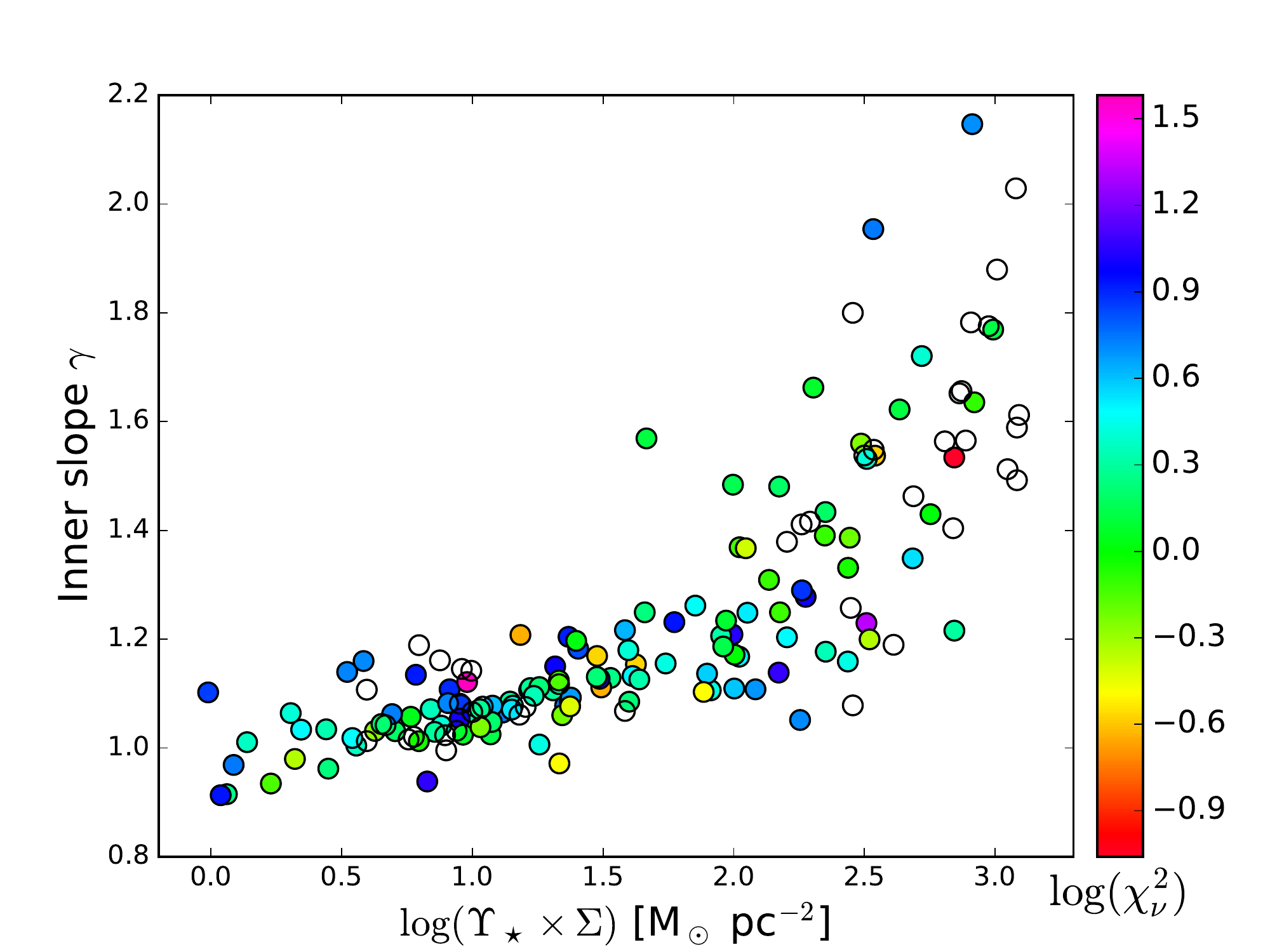}\includegraphics[scale=0.4]{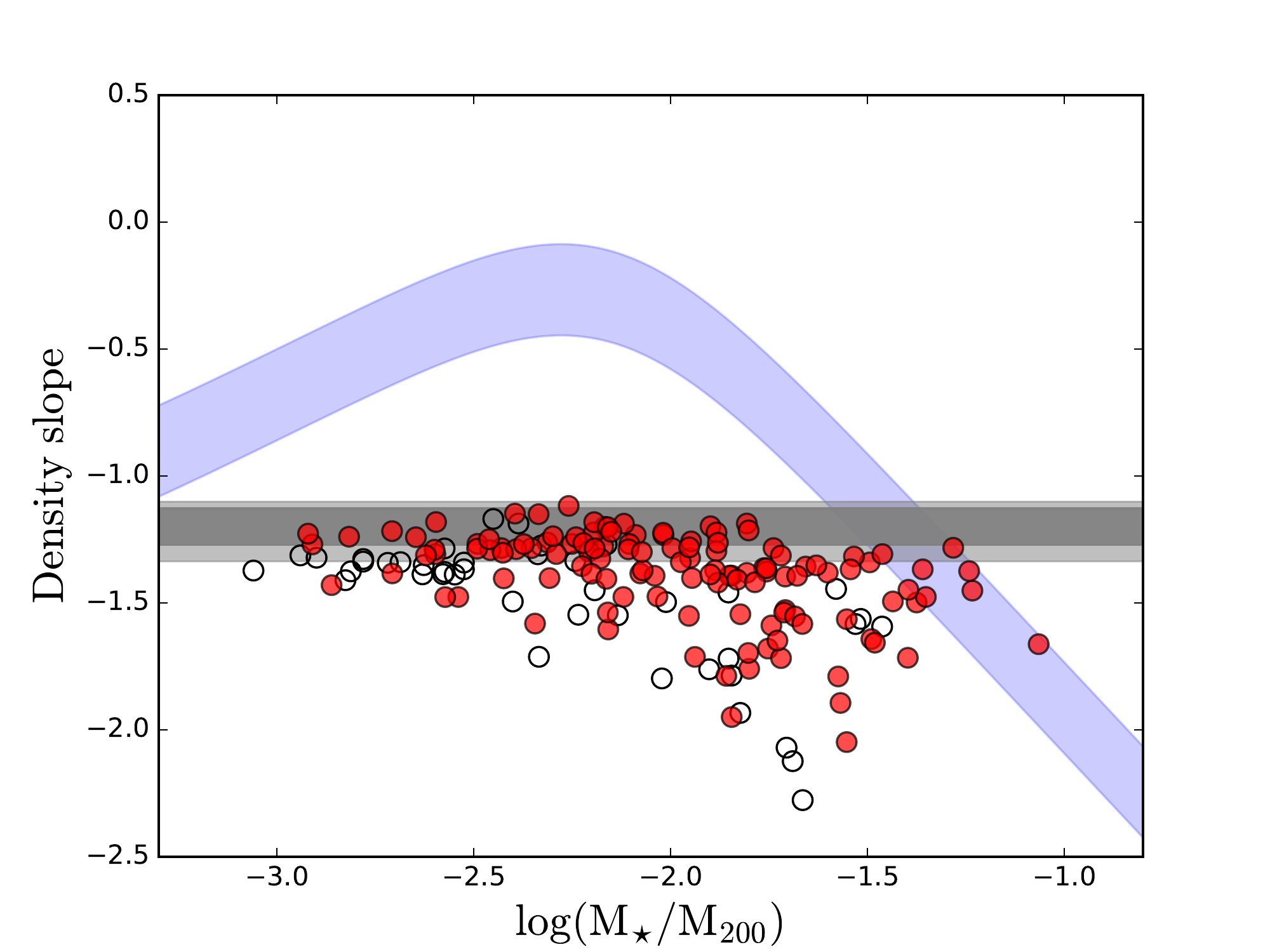}
    \caption{Inner density slopes of compressed halos. Left: The best-fit inner slopes of the compressed DM halo profiles versus effective surface mass density. For galaxies with a bulge, the value of $\Upsilon_\star$ is chosen as $\Upsilon_{\rm bulge}$. Filled points show galaxies that are fit by compressed halos, color coded by their fit qualities. Open points are galaxies for which our fitting procedure fails, so we compressed initial NFW halos determined by $\Lambda$CDM priors. Right: the density slopes measured at 1.5\% $R_{200}$ plotted against stellar-to-halo mass ratio. The blue region shows the density slope within 1 $\sigma$ scatter expected from hydrodynamic simulations of galaxy formation with strong stellar feedback \citep{Tollet2016}. Dark grey and light grey regions show the expected slopes of NFW halos for the SPARC galaxies according to their concentrations with and without considering the 1 $\sigma$ scatter, respectively. Open points have the same meaning as in the left panel.}
    \label{fig:Gamma}
\end{figure*}

\subsection{Galaxies that fail to run compression}

The current version of the compression code cannot accommodate the big diversity of the SPARC galaxies. There are 44 galaxies out of 169 failing to fit their rotation curves with compressed halos. We plot the effective surface brightnesses and the luminosities of these failed galaxies together with those successful ones in Figure \ref{fig:GalaxyFail}. Interestingly, the failed galaxies are distributed at two extremes: either being the most or the least luminous.

In order to investigate how baryonic compression affects those failed galaxies, we simply set their stellar mass-to-light ratios as the fiducial values ($\Upsilon_{\rm disk}=0.5$, $\Upsilon_{\rm bulge}=0.7$). We assigned them the NFW halos, and determined their halo masses and concentrations according to the abundance matching by \citet{Moster2013} and the halo mass-concentration relation by \citet{DuttonMaccio2014}, respectively. These halos are simply $\Lambda$CDM priors. We then compressed these prior halos and investigated how their rotation curves are compared with observations. Since we chose an intermediate stellar mass-to-light ratio, most of the galaxies are avoided from the two extremes where galaxies are easy to fail (except for the two least luminous galaxies that fail to run compression nevertheless). With fixed initial parameters, we did not need to scan the parameter space. This dramatically reduces the probability that any unexpected coincidence can happen. As a result, we were able to obtain the compressed halos for 42 galaxies among the 44 failed ones.

Figure \ref{fig:FailExample} shows the rotation curves of NGC 2903 as an example. This galaxy has a high central surface brightness and its disk is nearly maximal. This leaves no much room for dark matter. As a result, the DM halos is strongly compressed and the resulting rotation curve well overshoot the data. This is a general behaviour for high surface-brightness galaxies in the $\Lambda$CDM framework. In order to achieve a better fit, one has to reduce the stellar mass-to-light ratio, and perhaps also the concentration of the initial NFW halo.

\subsection{Properties of compressed halos}

After adiabatic compression, DM halo profiles are not described anymore by the NFW model, but they can be well fitted using the ($\alpha$, $\beta$, $\gamma$) model with $\alpha=1$. In the left panel of Figure \ref{fig:Gamma}, we plot the best-fit inner slope $\gamma$ against effective surface mass density. The effective surface mass density is calculated using their best-fit $\Upsilon_{\rm disk}$ for disk-only galaxies and $\Upsilon_{\rm bulge}$ for bulge-dominated galaxies. We color code galaxies with the reduced $\chi^2$ to indicate the qualities of their rotation curve fits.

The best-fit values of $\gamma$ show a strong correlation with surface mass density: high surface-mass-density galaxies tend to have denser DM halos towards their centers. When the surface mass density is low, the adiabatic contraction is so insignificant that the original NFW halo persists. As a result, an inner slope of one is recovered. For high surface-mass-density galaxies, the inner slope could be as high as two, which doubles the initial NFW slope. This implies that the baryonic effects could make DM halos even more cuspy than the self-gravity of the DM halo itself. We also plot the inner slopes of the compressed prior halos for the 42 failed galaxies, and find they follow the same trend.

We also investigated the density slopes at 1.5\% $R_{\rm 200}$ which are commonly used in cosmological simulations \citep[e.g.][]{Tollet2016}. The right panel of Figure \ref{fig:Gamma} plots the measured slopes together with the expectations of the NFW model. The NFW density slopes at a certain percentage $x$ of $R_{200}$ are given by:
\begin{equation}
    \frac{{\rm d}\log\rho}{{\rm d}\log r}= -\frac{1+3xC_{200}}{1+xC_{200}}.
\end{equation}
For example, for $x=0$ (the center of the halo) one recovers the inner slope of 1, while for $x\rightarrow\infty$ one recovers the outer slope of 3. In our specific case, $x=1.5\%$. Since such inner slope depends on halo concentration, Figure\,\ref{fig:Gamma} shows the expected slopes as a band using the concentration range spanned by the SPARC galaxies: ($C_{200}(M_{\rm 200,min})$, $C_{200}(M_{\rm 200,max})$) according to the mean halo mass-concentration relation from \citet{DuttonMaccio2014}. To take into account the scatter on this relation, we draw a lightly shaded region using the concentration range ($C_{200}(M_{\rm 200,max})$ - 1 $\sigma$, $C_{200}(M_{\rm 200,min})$) + 1 $\sigma$). The expected slopes are much less steeper than that of the vast majority of the compressed halos. Only when the stellar-to-halo mass ratios are low, the compressed halos are close to NFW. For galaxies with high stellar-to-halo mass ratios, the slopes of the compressed halos significantly differ from the NFW model. As a reference, we also plot the expected range of slopes derived from hydrodynamic simulations of galaxy formation with strong stellar feedback \citep{Tollet2016}. Current simulations have implemented various feedback mechanisms that are sufficiently strong to turn the grey band into the blue band. Our results show that adiabatically compressed DM halos are more cuspy than NFW halos even for relatively small values of $M_\star/M_{200}\simeq-2.0$, so baryonic feedback should not simply turn a NFW cusp into a core but also counteract the important effect of baryon-driven halo compression.

\section{Discussion and conclusions}

In this paper, we studied the baryon-driven compression of DM halos in disk galaxies from the SPARC sample. We first illustrated the magnitude of baryonic compression by evolving the best-fit fixed NFW halos that are derived with traditional rotation-curve fits. We find that high-surface-brightness galaxies (with $\Sigma_{\rm eff}\gtrsim100$ L$_\odot$ pc$^{-2}$ at 3.6 $\mu$m) must experience a strong baryonic compression, so that their compressed halos are significantly different from the initial NFW halos. This implies that the best-fit fixed NFW halos are not in dynamical equilibrium with the embedded baryons. Therefore, the traditional approach fails to derive reasonable NFW halos by fitting rotation curves.

We developed a new rotation-curve fitting methodology that takes adiabatic compression into account while searching for the best-fit masses and concentrations of the initial NFW halos (before compression). This guarantees the dynamical stability of the final system. The compressed halos do not retain the NFW profile, but generally have steeper density profiles. Their inner slopes systematically increase with the baryonic surface mass density. However, to obtain satisfactory fits, one has to systematically tune down stellar mass-to-light ratios and halo concentrations. This may indicate that massive galaxies suffer a similar cusp-core problem as dwarf galaxies: to make room for a central DM cusp, the stellar contributions must be systematically lower than expected from stellar population synthesis models.

In this study, we intentionally isolated and quantified the baryonic gravitational effects (adiabatic halo contraction) from that of feedback (possibly leading to halo expansion). Stellar feedback is believed to be a plausible solution to the core-cusp problem within the $\Lambda$CDM framework, but it is designed to work in dwarf galaxies and so is rather ineffective in massive galaxies. Thus, existing feedback mechanisms may reduce the discrepancy for some galaxies in our sample, but are expected to be ineffective for the most massive galaxies. Our results suggest that the most massive galaxies experience the strongest adiabatic contractions, so some new form of feedback mechanisms that can work efficiently in massive galaxies is required to counteract the halo contraction in a $\Lambda$CDM context. One possibility may be feedback from active galactic nuclei (AGN).

The same problem has been identified in  \citet{Li2022} by building model galaxies within the $\Lambda$CDM framework. In that study, we calculated the radial acceleration relation \citep[RAR,][]{McGaugh2016PRL, OneLaw} for each individual model galaxy. The modeled RAR for low-mass galaxies present systematically upward ``hooks'', suggesting a core-cusp problem, while high-mass galaxies are consistent with the observed RAR. However, once we include adiabatic contraction, the RAR of massive galaxies systematically shifts above the observed RAR. This implies that high-mass galaxies suffer a similar core-cusp problem as dwarf galaxies. In a $\Lambda$CDM context, therefore, feedback processes must be fine tuned because they must precisely compensate the shift from the observed RAR due to adiabatic contraction in order to move model galaxies back to the observed relation.

In both studies, we find that the strength of adiabatic contraction is correlated with baryonic surface mass density, so any plausible feedback mechanisms must have a strength that correlates with baryonic surface mass density in order to precisely counteract the effect of adiabatic contraction for all galaxies.

\begin{acknowledgements}
P.L. is supported by the Alexander von Humboldt Foundation. S.S.M. and J.M.S. are supported in part by NASA ADAP grant 80NSSC19k0570. S.S.M. also acknowledges support from NSF PHY-1911909. M.S.P. acknowledges funding of a Leibniz-Junior Research Group (project number J94/2020) and a KT Boost Fund by the German Scholars Organization and Klaus Tschira Stiftung. 
\end{acknowledgements}

\bibliographystyle{aa}
\bibliography{PLi}

\longtab[1]{
\begin{longtable}{lccccccccc}
\caption{Best-fit parameters for stellar disks, bulges, primordial NFW halos and compressed halos. $V_{200}$ and $C_{200}$ are for the primordial NFW halos; $\rho_s$, $\gamma$ and $\beta$ are the parameters of the compressed halos in the ($\alpha$, $\beta$, $\gamma$) models, where the transition parameter has been fixed at $\alpha=1$; $r_s$ is the shared scale length of the prior-compression and post-compression halos given it is the unit in which the adiabatic contraction is calculated.}
\label{tab:parameters}\\
\hline\hline
{Galaxy name} & {$\Upsilon_{\rm disk}$} & {$\Upsilon_{\rm bulge}$} & {$V_{200}$} & {$C_{200}$} & {$r_s$} & {$\log(\rho_s)$} & {$\gamma$} & {$\beta$} & $\chi^2_\nu$ \\
 {}& {($M_\odot/L_\odot$)} & {($M_\odot/L_\odot$)} & {(km/s)} & {} & {(kpc)} & {[$M_\odot$ kpc$^{-3}$]} & {} & {} \\
\hline
\endfirsthead
\caption{continued}\\
\hline
{Galaxy name} & {$\Upsilon_{\rm disk}$} & {$\Upsilon_{\rm bulge}$} & {$V_{200}$} & {$C_{200}$} & {$r_s$} & {$\log(\rho_s)$} & {$\gamma$} & {$\beta$} & $\chi^2_\nu$\\
 {}& {($M_\odot/L_\odot$)} & {($M_\odot/L_\odot$)} & {(km/s)} & {} & {(kpc)} & {[$M_\odot$ kpc$^{-3}$]} & {} & {} \\
\hline
\endhead
\hline
\endfoot
\hline
\endlastfoot
DDO161 & 0.16 & \dots & 60.70 & 4.35 & 19.11 & 5.97 & 1.14 & 3.45 & 5.31\\
DDO170 & 0.13 & \dots & 48.34 & 7.54 & 8.78 & 6.64 & 0.97 & 3.37 & 5.52\\
ESO079-G014 & 0.77 & \dots & 214.72 & 3.38 & 87.14 & 5.65 & 1.11 & 3.11 & 4.94\\
ESO116-G012 & 0.25 & \dots & 88.09 & 10.59 & 11.39 & 6.91 & 1.08 & 3.22 & 5.46\\
ESO563-G021 & 0.75 & \dots & 447.78 & 3.08 & 199.13 & 5.34 & 1.23 & 2.15 & 21.58\\
F561-1 & 0.10 & \dots & 57.28 & 3.57 & 21.97 & 5.91 & 1.03 & 3.86 & 3.12\\
F563-1 & 0.98 & \dots & 80.53 & 8.45 & 13.06 & 6.67 & 1.09 & 3.18 & 2.02\\
F563-V1 & 0.10 & \dots & 49.34 & 2.52 & 26.82 & 5.47 & 1.10 & 3.90 & 7.19\\
F563-V2 & 1.00 & \dots & 81.58 & 8.97 & 12.46 & 6.81 & 1.08 & 3.48 & 2.13\\
F565-V2 & 1.00 & \dots & 63.84 & 8.40 & 10.41 & 6.66 & 1.07 & 3.17 & 2.28\\
F567-2 & 0.10 & \dots & 52.53 & 5.85 & 12.30 & 6.44 & 0.92 & 3.47 & 1.85\\
F568-1 & 1.00 & \dots & 96.21 & 8.48 & 15.54 & 6.72 & 1.11 & 3.29 & 1.98\\
F568-3 & 1.00 & \dots & 118.81 & 2.26 & 71.91 & 5.34 & 1.09 & 4.00 & 4.72\\
F568-V1 & 0.98 & \dots & 84.43 & 10.26 & 11.27 & 6.89 & 1.13 & 3.21 & 0.27\\
F571-8 & 0.18 & \dots & 131.47 & 7.54 & 23.90 & 6.43 & 1.14 & 2.90 & 11.71\\
F571-V1 & 0.38 & \dots & 65.47 & 7.76 & 11.56 & 6.61 & 1.06 & 3.29 & 1.12\\
F574-1 & 0.55 & \dots & 85.18 & 6.28 & 18.57 & 6.40 & 1.11 & 3.35 & 2.42\\
F574-2 & 0.10 & \dots & 54.72 & 3.29 & 22.79 & 5.93 & 0.91 & 4.11 & 8.72\\
F579-V1 & 0.39 & \dots & 83.32 & 11.33 & 10.08 & 7.05 & 1.06 & 3.27 & 0.61\\
F583-1 & 0.68 & \dots & 69.06 & 5.82 & 16.25 & 6.33 & 1.04 & 3.39 & 2.57\\
F583-4 & 0.17 & \dots & 57.31 & 8.16 & 9.62 & 6.65 & 1.03 & 3.24 & 0.52\\
NGC0024 & 0.90 & \dots & 84.77 & 8.20 & 14.17 & 6.55 & 1.31 & 3.25 & 0.79\\
NGC0055 & 0.15 & \dots & 72.95 & 5.67 & 17.62 & 6.26 & 1.11 & 3.38 & 7.81\\
NGC0100 & 0.22 & \dots & 71.13 & 7.96 & 12.24 & 6.61 & 1.08 & 3.21 & 2.61\\
NGC0247 & 1.00 & \dots & 120.71 & 2.83 & 58.47 & 5.51 & 1.13 & 3.68 & 1.82\\
NGC0289 & 0.44 & \dots & 133.15 & 5.66 & 32.20 & 6.30 & 1.22 & 3.45 & 2.02\\
NGC0300 & 0.27 & \dots & 78.46 & 8.94 & 12.02 & 6.70 & 1.09 & 3.14 & 1.66\\
NGC1003 & 0.58 & \dots & 108.99 & 3.77 & 39.57 & 5.85 & 1.11 & 3.53 & 2.65\\
NGC1090 & 0.37 & \dots & 124.18 & 5.89 & 28.89 & 6.39 & 1.17 & 3.65 & 3.05\\
NGC1705 & 1.00 & \dots & 62.22 & 10.30 & 8.28 & 6.44 & 1.54 & 2.12 & 0.26\\
NGC2403 & 0.29 & \dots & 99.56 & 10.38 & 13.13 & 6.82 & 1.21 & 3.07 & 10.92\\
NGC2683 & 0.46 & 0.20 & 109.07 & 8.19 & 18.24 & 6.40 & 1.72 & 2.93 & 2.54\\
NGC2841 & 1.00 & 0.49 & 330.63 & 3.28 & 138.21 & 4.91 & 1.77 & 0.87 & 1.34\\
NGC2915 & 0.29 & \dots & 58.67 & 15.77 & 5.10 & 7.23 & 1.17 & 2.91 & 1.02\\
NGC2976 & 0.73 & \dots & 81.44 & 1.58 & 70.47 & 5.16 & 1.18 & 10.97 & 2.24\\
NGC2998 & 0.57 & \dots & 172.89 & 4.15 & 57.10 & 5.98 & 1.16 & 3.53 & 2.76\\
NGC3198 & 0.50 & \dots & 123.29 & 5.35 & 31.56 & 6.19 & 1.21 & 3.37 & 2.06\\
NGC3726 & 0.36 & \dots & 139.52 & 4.29 & 44.53 & 5.90 & 1.26 & 3.32 & 2.97\\
NGC3769 & 0.24 & \dots & 89.60 & 9.18 & 13.37 & 6.71 & 1.25 & 3.14 & 0.76\\
NGC3877 & 0.30 & \dots & 133.60 & 6.57 & 27.85 & 6.67 & 1.05 & 4.71 & 5.11\\
NGC3893 & 0.36 & \dots & 133.12 & 8.16 & 22.34 & 6.46 & 1.43 & 2.98 & 1.05\\
NGC3917 & 0.89 & \dots & 157.22 & 2.82 & 76.38 & 5.54 & 1.11 & 3.85 & 3.92\\
NGC3949 & 0.29 & \dots & 102.30 & 8.41 & 16.66 & 6.43 & 1.56 & 2.95 & 0.56\\
NGC3953 & 0.38 & \dots & 155.87 & 6.69 & 31.94 & 6.36 & 1.39 & 3.44 & 0.79\\
NGC3972 & 0.48 & \dots & 101.03 & 7.03 & 19.68 & 6.57 & 1.14 & 3.88 & 3.79\\
NGC3992 & 0.76 & \dots & 254.00 & 2.79 & 124.68 & 5.30 & 1.33 & 2.94 & 0.91\\
NGC4010 & 0.26 & \dots & 96.48 & 7.60 & 17.40 & 6.57 & 1.16 & 3.38 & 5.13\\
NGC4013 & 0.20 & 1.00 & 146.62 & 4.81 & 41.78 & 5.86 & 1.48 & 2.91 & 1.58\\
NGC4085 & 0.22 & \dots & 91.73 & 9.58 & 13.12 & 6.76 & 1.28 & 3.40 & 10.04\\
NGC4088 & 0.23 & \dots & 124.44 & 5.88 & 29.01 & 6.23 & 1.37 & 3.35 & 0.75\\
NGC4100 & 0.52 & \dots & 118.39 & 7.02 & 23.11 & 6.21 & 1.66 & 2.86 & 1.20\\
NGC4157 & 0.39 & 0.10 & 154.57 & 4.95 & 42.78 & 6.11 & 1.20 & 3.43 & 0.45\\
NGC4183 & 0.36 & \dots & 80.52 & 8.66 & 12.73 & 6.75 & 1.11 & 3.33 & 0.23\\
NGC4217 & 1.00 & 0.10 & 153.42 & 7.96 & 26.39 & 6.45 & 1.35 & 3.04 & 3.50\\
NGC4389 & 0.10 & \dots & 77.98 & 6.55 & 16.31 & 6.46 & 1.15 & 3.99 & 9.85\\
NGC4559 & 0.20 & \dots & 88.53 & 8.69 & 13.95 & 6.72 & 1.15 & 3.25 & 0.27\\
NGC5005 & 0.39 & 0.39 & 239.94 & 6.07 & 54.17 & 5.93 & 1.53 & 2.90 & 0.09\\
NGC5585 & 0.10 & \dots & 71.62 & 8.48 & 11.58 & 6.66 & 1.08 & 3.16 & 7.25\\
NGC5985 & 0.19 & 0.88 & 189.24 & 17.65 & 14.69 & 7.44 & 1.16 & 3.09 & 2.69\\
NGC6015 & 0.55 & \dots & 114.75 & 6.74 & 23.33 & 6.42 & 1.29 & 3.36 & 7.42\\
NGC6503 & 0.29 & \dots & 87.13 & 10.01 & 11.92 & 6.64 & 1.43 & 2.84 & 1.58\\
NGC6674 & 0.61 & 0.97 & 282.57 & 2.15 & 179.68 & 4.37 & 1.95 & 0.99 & 5.46\\
NGC6789 & 1.00 & \dots & 48.95 & 11.63 & 5.76 & 6.88 & 1.23 & 3.88 & 8.75\\
NGC7793 & 0.40 & \dots & 83.80 & 7.03 & 16.32 & 6.48 & 1.23 & 3.81 & 1.24\\
NGC7814 & 0.93 & 0.33 & 166.31 & 7.22 & 31.55 & 5.91 & 1.84 & 1.77 & 0.50\\
UGC00128 & 0.33 & \dots & 98.38 & 8.73 & 15.44 & 6.83 & 0.94 & 3.40 & 11.47\\
UGC00191 & 0.26 & \dots & 62.71 & 9.15 & 9.39 & 6.79 & 1.07 & 3.28 & 5.50\\
UGC00634 & 0.45 & \dots & 81.44 & 8.56 & 13.03 & 6.72 & 1.07 & 3.26 & 23.03\\
UGC00731 & 0.84 & \dots & 55.33 & 7.99 & 9.48 & 6.73 & 0.97 & 3.42 & 0.33\\
UGC00891 & 0.50 & \dots & 54.05 & 6.87 & 10.77 & 6.42 & 1.12 & 3.22 & 38.20\\
UGC01230 & 0.41 & \dots & 80.13 & 8.50 & 12.91 & 6.78 & 1.02 & 3.37 & 1.35\\
UGC01281 & 0.26 & \dots & 50.36 & 6.49 & 10.63 & 6.47 & 1.00 & 3.62 & 2.40\\
UGC02023 & 0.10 & \dots & 50.27 & 7.29 & 9.45 & 6.57 & 1.03 & 3.46 & 2.20\\
UGC02259 & 0.97 & \dots & 70.85 & 8.26 & 11.74 & 6.56 & 1.25 & 3.16 & 1.71\\
UGC02455 & 0.10 & \dots & 57.72 & 2.89 & 27.31 & 5.65 & 1.18 & 5.60 & 6.63\\
UGC02487 & 1.00 & 0.83 & 454.69 & 1.44 & 432.12 & 3.60 & 2.15 & -0.93 & 5.07\\
UGC02885 & 0.11 & 0.69 & 254.34 & 6.39 & 54.53 & 5.99 & 1.57 & 2.46 & 1.32\\
UGC03205 & 0.55 & 0.77 & 173.22 & 4.53 & 52.38 & 5.76 & 1.53 & 2.82 & 2.57\\
UGC03580 & 0.26 & 0.11 & 103.55 & 6.74 & 21.06 & 6.35 & 1.20 & 3.11 & 3.11\\
UGC04278 & 0.35 & \dots & 68.67 & 7.20 & 13.06 & 6.50 & 1.08 & 3.20 & 4.12\\
UGC04325 & 1.00 & \dots & 72.96 & 8.82 & 11.33 & 6.79 & 1.13 & 3.75 & 3.22\\
UGC04499 & 0.15 & \dots & 56.49 & 8.88 & 8.71 & 6.79 & 1.03 & 3.33 & 1.55\\
UGC05005 & 0.28 & \dots & 72.21 & 5.98 & 16.54 & 6.38 & 1.03 & 3.41 & 2.00\\
UGC05414 & 0.15 & \dots & 53.33 & 7.34 & 9.96 & 6.56 & 1.06 & 3.44 & 5.43\\
UGC05716 & 0.35 & \dots & 58.50 & 8.14 & 9.85 & 6.66 & 1.04 & 3.25 & 2.33\\
UGC05721 & 0.39 & \dots & 52.37 & 19.84 & 3.62 & 7.51 & 1.19 & 2.97 & 1.39\\
UGC05750 & 0.20 & \dots & 65.04 & 5.05 & 17.65 & 6.19 & 1.01 & 3.44 & 2.34\\
UGC05829 & 0.16 & \dots & 48.15 & 6.82 & 9.67 & 6.62 & 0.93 & 3.64 & 0.72\\
UGC05918 & 0.10 & \dots & 39.35 & 8.06 & 6.68 & 6.69 & 0.98 & 3.34 & 0.24\\
UGC05986 & 0.40 & \dots & 97.83 & 9.28 & 14.44 & 6.75 & 1.13 & 3.21 & 10.60\\
UGC05999 & 0.39 & \dots & 74.29 & 6.92 & 14.70 & 6.51 & 1.05 & 3.35 & 10.26\\
UGC06399 & 0.54 & \dots & 71.57 & 8.07 & 12.16 & 6.62 & 1.11 & 3.27 & 1.95\\
UGC06446 & 0.81 & \dots & 65.09 & 8.76 & 10.18 & 6.67 & 1.17 & 3.13 & 0.27\\
UGC06628 & 0.10 & \dots & 56.82 & 4.11 & 18.93 & 6.08 & 1.02 & 3.96 & 3.07\\
UGC06667 & 1.00 & \dots & 72.99 & 8.34 & 11.99 & 6.72 & 1.01 & 3.35 & 2.69\\
UGC06786 & 0.28 & 0.40 & 156.49 & 13.31 & 16.11 & 6.99 & 1.39 & 2.91 & 0.62\\
UGC06917 & 0.34 & \dots & 82.52 & 9.02 & 12.54 & 6.74 & 1.11 & 3.22 & 1.68\\
UGC06923 & 0.18 & \dots & 61.94 & 10.11 & 8.39 & 6.85 & 1.13 & 3.22 & 1.67\\
UGC06930 & 0.32 & \dots & 80.03 & 8.38 & 13.08 & 6.72 & 1.08 & 3.34 & 0.37\\
UGC06973 & 0.13 & 0.23 & 128.14 & 12.43 & 14.13 & 6.64 & 1.62 & 2.13 & 1.31\\
UGC06983 & 0.40 & \dots & 79.61 & 10.28 & 10.61 & 6.88 & 1.12 & 3.18 & 0.72\\
UGC07089 & 0.12 & \dots & 61.68 & 6.59 & 12.82 & 6.45 & 1.04 & 3.36 & 1.63\\
UGC07125 & 0.10 & \dots & 49.72 & 4.38 & 15.53 & 6.23 & 0.96 & 3.78 & 1.73\\
UGC07151 & 0.50 & \dots & 68.13 & 6.23 & 14.98 & 6.31 & 1.22 & 3.46 & 4.28\\
UGC07232 & 0.28 & \dots & 41.62 & 12.20 & 4.67 & 6.89 & 1.20 & 2.76 & 8.35\\
UGC07261 & 0.39 & \dots & 63.21 & 8.28 & 10.46 & 6.58 & 1.21 & 3.09 & 0.23\\
UGC07323 & 0.23 & \dots & 70.53 & 6.93 & 13.93 & 6.52 & 1.07 & 3.57 & 3.46\\
UGC07399 & 0.99 & \dots & 77.94 & 12.28 & 8.69 & 6.92 & 1.25 & 2.83 & 3.36\\
UGC07524 & 0.31 & \dots & 69.47 & 6.20 & 15.35 & 6.42 & 1.02 & 3.45 & 1.21\\
UGC07603 & 0.21 & \dots & 50.15 & 12.50 & 5.50 & 7.05 & 1.10 & 3.08 & 2.25\\
UGC07608 & 0.72 & \dots & 54.18 & 8.68 & 8.55 & 6.73 & 1.05 & 3.30 & 1.58\\
UGC07690 & 0.54 & \dots & 59.43 & 6.21 & 13.11 & 5.96 & 1.48 & 2.15 & 1.41\\
UGC08286 & 1.00 & \dots & 67.84 & 8.22 & 11.31 & 6.62 & 1.18 & 3.33 & 2.60\\
UGC08490 & 0.89 & \dots & 63.43 & 8.82 & 9.85 & 6.52 & 1.37 & 2.84 & 0.40\\
UGC08550 & 0.55 & \dots & 50.96 & 8.60 & 8.12 & 6.57 & 1.20 & 2.94 & 1.06\\
UGC08699 & 0.38 & 0.46 & 133.09 & 7.83 & 23.27 & 6.27 & 1.64 & 2.70 & 0.80\\
UGC09037 & 0.13 & \dots & 128.52 & 5.61 & 31.36 & 6.27 & 1.13 & 3.37 & 2.14\\
UGC09992 & 0.10 & \dots & 41.04 & 6.25 & 9.00 & 6.39 & 1.06 & 3.43 & 2.58\\
UGC10310 & 0.22 & \dots & 58.49 & 8.87 & 9.04 & 6.80 & 1.01 & 3.37 & 0.91\\
UGC11557 & 0.10 & \dots & 69.84 & 5.61 & 17.07 & 6.33 & 1.07 & 3.57 & 1.95\\
UGC11820 & 0.40 & \dots & 70.61 & 5.52 & 17.51 & 6.24 & 1.08 & 3.33 & 5.24\\
UGC12506 & 0.53 & \dots & 159.40 & 10.38 & 21.04 & 7.00 & 1.10 & 3.36 & 0.33\\
UGC12632 & 0.16 & \dots & 53.61 & 9.27 & 7.92 & 6.85 & 0.98 & 3.28 & 0.46\\
UGC12732 & 0.40 & \dots & 68.84 & 7.65 & 12.33 & 6.62 & 1.04 & 3.28 & 0.58\\
UGCA442 & 0.80 & \dots & 52.78 & 6.32 & 11.44 & 6.32 & 1.13 & 3.32 & 8.39\\
\end{longtable}
}

\end{document}